\documentclass{aa}

\usepackage{graphicx}
\usepackage{txfonts}
\usepackage{natbib}
\usepackage{amsfonts}
\usepackage{amssymb}

\usepackage{amstext}

\begin{document}

\title{Construction of coronal hole and active region  magnetohydrostatic solutions in two dimensions: \\ 
Force and energy balance}

\author{J. Terradas\inst{1,2}, R. Soler\inst{1,2}, R. Oliver\inst{1,2}, P. Antolin\inst{3}, I. Arregui\inst{4,5},\\ 
M. Luna\inst{1,2}, I. Piantschitsch\inst{1,2,6}, E. Soubri\'e\inst{2,7}, and J. L. Ballester\inst{1,2}}
\institute{$^1$Departament de F\'\i sica, Universitat de les Illes Balears (UIB),
E-07122, Spain \\ $^2$Institute of Applied Computing \& Community Code (IAC$^3$),
UIB, Spain\\ \email{jaume.terradas@uib.es}\\
$^3$Department of Mathematics, Physics and Electrical Engineering, Northumbria University, Newcastle Upon Tyne,
NE1 8ST, UK\\
$^4$Instituto de Astrof\'\i sica de Canarias, V\'\i a L\'actea s/n, 38205 La Laguna, Tenerife, Spain\\
$^5$Departamento de Astrof\'\i sica Universidad de La Laguna, 38206 La Laguna, Tenerife, Spain\\
$^6$Institute of Physics, University of Graz, Universitätsplatz 5, 8010 Graz, Austria\\
$^7$Institut d'Astrophysique Spatiale, CNRS, Univ. Paris-Sud, Université Paris-Saclay, Bât. 121, 91405 Orsay cedex, France
}
\abstract{
Coronal holes and active regions are typical magnetic structures found in the solar atmosphere. We propose several magnetohydrostatic equilibrium solutions that are representative of these  structures in two dimensions. Our models include the effect of a finite plasma-$\beta$ and gravity, but the distinctive feature is that we incorporate a thermal structure with properties similar to those reported by observations. We developed a semi-analytical method to compute the equilibrium configuration. Using this method, we obtain cold and under-dense plasma structures in open magnetic fields representing coronal holes, while in closed magnetic configurations, we achieve the characteristic hot and over-dense plasma arrangements of active regions. Although coronal holes and active regions seem to be antagonistic structures, we find that they can be described using a common thermal structure that depends on the flux function. In addition to the force balance, the energy balance is included in the constructed models using an a posteriori approach. From the two-dimensional computation of thermal conduction and radiative losses in our models, we infer the required heating function to achieve energy equilibrium.  We find that the temperature dependence on height is an important parameter that may prevent the system from accomplishing thermal balance at certain spatial locations. The implications of these results are discussed in detail.
}

\date{\today}

\titlerunning{Coronal hole and active region MHS solutions in 2D}
\authorrunning{Terradas et al.}
\maketitle

\section{Introduction}

Coronal holes (CHs) are usually defined as the darkest patches on
the solar surface that are observed in ultraviolet (UV) and
X-ray radiation. These structures are associated with magnetic fields that are open to the interplanetary space and have strong links with the solar wind, which is thought to emanate from their base. A significant effort over the past decades of part of the scientific solar community has been to reproduce the velocities that are reported for the solar wind. We refer to the review of \citet{cranmer2009} of measurements of the plasma properties in CHs and how they are used
to reveal details about the physical processes that heat the solar corona and accelerate the solar
wind. Because the magnetic field in CHs is unipolar and therefore open, this configuration is in some aspects similar to solar sunspots \citep[see a detailed comparison in][]{obridkosolovev2011}, which are characterised by a concentration of magnetic flux  at a low temperature with respect to the environment. However, CHs are thought to have a low plasma-$\beta$ (ratio of gas to magnetic pressure) in contrast to sunspots, where this parameter changes substantially with height from the photosphere to the corona. The plasma-$\beta$ is typically about $3\times 10^{-3}$ in CHs \citep[see e.g.][]{isenvas2007}.

The arrangement of active regions (ARs) is inverse to that of CHs. ARs are composed of dense and hot plasma cores lying along closed magnetic field lines. UV-imaging spectroscopy during the early
1970s from Skylab already revealed that ARs are composed of filamentary structures, commonly called loops, rather than consisting of a plain diffuse plasma distribution. \citet{gallagher2001} found that the AR they studied had the structure of a central
hot high-density core, contained by a large number of
low-lying magnetic field lines and surrounded by cooler loops, which in turn were embedded in hot coronal plasma. The authors were able to estimate the plasma pressure distribution over the AR and found that it was higher in the core and lower
in the halo (almost two orders of magnitude). This agrees with the fact that pressure in ARs is known to be correlated with higher magnetic flux
concentrations; see for example \citet{golubetal1980}. It is also known that the magnetic field is a primary quantity in relation to coronal heating, especially in ARs. Although the analysis and modelling of coronal loops in ARs has received much attention \citep[see e.g.][]{aschwandenetal1999a,aschwandenetal2000a}, the plasma emitting in the EUV  within the AR that is not associated with distinguishable loops is often ignored. Here we concentrate on the largely diffuse component in ARs because coronal loops generally comprise only
a small proportion of the AR emission, typically only 10\%–30\% of  the enhancements over the background (\citealt{Delzannamason2003} and \citealt{viallklimchuk2011}). In our case, and similar to the definition given in \citet{viallklimchuk2011}, diffuse emission refers to background emission without distinct  discrete intensity enhancement such as loops, and whether it occurs in hot or warm temperatures. One of the conclusions of \citet{viallklimchuk2011} was that the diffuse background shows understandable patterns that are consistent with impulsive nanoflare heating. This indicates that the emitting coronal plasma in the whole AR is not steady, but dynamic and constantly evolving \citep[see also][]{viallklimchuk2012,viallklimchuk2016,viallklimchuk2017}. This somewhat contradicts the results of \citet{warrenetal2012}, who indicated that the hot-component plasma in
several observed ARs is often close to equilibrium. Recently, \citet{warrenetal2020} have shown using observations of an AR from the High-resolution Coronal Imager (Hi-C) sounding rocket together with  modelling that the heating in the AR core is taking place at a relatively high frequency and
that the observed loops are close to equilibrium. This is an indication of nearly steady heating.

An approach that is commonly used in the literature for heating studies in the solar corona relies on the assumption of an a priori form of the heating, for example, a decaying exponential with height. This heating profile, representing a footpoint heating, has been used to solve the time-dependence problem, and the characteristics of the equilibrium, if achieved, or of the non-equilibrium, have been investigated in detail in the past (see e.g. the summaries of this subject of \citealt{klimchuk2019} and \citealt{antolin2020}). Thermal non-equilibrium (TNE) is a clear example of a non-existing equilibrium situation in which a non-linear mechanism can explain coronal rain or the cold condensations that lead to prominence formation around magnetic dips. The a priori approach to the heating form is not the only way to solve this problem. An alternative method is based on the calculation of the stationary solutions of the configuration in two or three dimensions. With this approach, we have the possibility of choosing a certain temperature and pressure profile, calculate the force balance to have equilibrium of forces, and investigate the energy balance a posterioir without an a priori assumption about the heating form. The choice of temperature and pressure, together with the magnetic field, determines thermal conduction in the system.  The optically thin radiative losses of the corona depend on the density and temperature distribution of the obtained solution under force balance. When we have the contribution of thermal conduction and radiative losses, we can calculate the heating profile to have a perfect equilibrium. An example of this is the one-dimensional (1D) Rosner-Tucker-Vaiana (RTV) scaling law (see the model of \citealt{rtv1978} under the assumption of uniform pressure and heating, and also the two-dimensional (2D) model of \citealt{petrieetal2003}). The question that arises here is whether we can always find energy balance in the system. With this approach we can derive, without directly solving the time-dependent problem, some conditions that lead to the existence or absence of thermal balance. In this case, thermal balance does not refer to thermal equilibrium along an individual magnetic field line only (the 1D case). Because the problem is treated in 2D in the present work, we instead consider CHs or ARs as global structures.

Regardless of whether the coronal heating is impulsive or steady, it is known that many magnetic structures of the solar corona  can from a large-scale point of view have long lifetimes. This suggests that they are in a sort of equilibrium state, although the observations reveal a strong dynamism that is often reinforced by flows on small spatial scales. In spite of the dynamic nature of the coronal structures, magnetohydrostatic (MHS) models, in which flows are assumed to be zero or stationary, should not be discarded. In particular, recent literature lacks MHS equilibrium models of CHs, except for \citet{tsinganos1981, tsinganos1982, lowtsinganos1986} in the 1980s, and the aforementioned work of \citet{obridkosolovev2011}. MHS models of CHs are needed for several reasons. They can provide a better understanding of how these structures are kept in the two basic balance conditions, namely the force and thermal balance. MHS models of CHs are necessary for other purposes such as for example to carry out investigations about the interaction of global MHD waves with these structures because so far, very simple geometries have been addressed, mostly based on a purely vertical magnetic field (see \citealt{piantschitschetal2018a,piantschitschetal2018b,piantschitschetal2020} and \citealt{piantschitschterradas2021}). Global three-dimensional (3D) MHD simulations have also been used to investigate this problem \citep[e.g.][]{downsetal2011}, but the results about the interaction with CHs are limited and deeper analyses are required, especially using elementary models. The same also applies to the construction of magnetohydrostatic solutions of ARs. Another argument in favour of studying MHS models is that we can obtain some information about the conditions for which these models are not possible. In order words, if the MHS models provide unrealistic configurations,  real solar structures are not in MHS equilibrium.

The main aim of the present work is twofold. First, we explore different CH models that include a cold and under-dense region (the CH) that connects with an atmosphere at typically 1 MK (corona) through a smooth coronal hole boundary (CHB). Using the same scheme based on the early works of \citet{low1975,low1980}, we investigate how uninvolved MHS models can also reproduce the main features of ARs, paying particular attention to the high pressure and diffuse background of these structures instead of the single-loop structures. The main characteristics of the different MHS models developed here are analysed in detail. Second, and closely related to the previous goal, we carry out a detailed investigation of the energy balance in these structures. We calculate the spatial distribution of heating required for a situation in thermal balance, and we provide some constraints for the temperature dependence on height to achieve energy equilibrium.

\section{Magnetohydrostatic equilibrium in 2D}

We start by describing the problem we aim to solve and how it is formulated using the equations of magnetohydrostatics, which are applicable as a first approach to CHs and ARs. 
\citet{low1975} addressed the situation of magnetohydrostatic equilibrium in the presence of purely vertical gravity in Cartesian geometry. We also refer to \citet{parker1968} (see their Appendix A) and \citet{parker1979}. We look for solutions to the following equation:
\begin{align}
    \frac{1}{\mu_0}(\nabla\times {\bf B})\times {\bf B}-\nabla p -\rho g \,{\bf {\hat e}_z}=0,\label{eq:mhdstatic}
\end{align}
where $\bf B$ is the magnetic field, $p$ is the gas pressure, $\rho$ is the plasma density, $g$ is the gravity acceleration on the solar surface (pointing in the negative $z-$direction), and $\mu_0$ is the magnetic permeability of free space. The magnetic field from Maxwell's equations has to satisfy
\begin{align}
\nabla \cdot {\bf B}=0.\label{eq:divb}
\end{align}
We assume that the plasma is composed of fully ionised hydrogen that satisfies the ideal gas law,
\begin{align}
p=\frac{\mathcal{R}}{\bar{\mu}} \rho  T,\label{eq:ideal}
\end{align}
where $T$ is the temperature, $\mathcal{R}$ is the gas constant, and $\bar{\mu}$ is the mean atomic weight ($\bar{\mu}=0.5$ for fully ionised hydrogen plasma and $\bar{\mu}=0.6$ when fully ionised helium with coronal abundances is included). The aim is to obtain solutions to the previous equations, but we have a system of five equations (Eqs.~(\ref{eq:mhdstatic})-(\ref{eq:ideal})) but six unknowns, $\bf B$ (three components), $p$, $\rho,$ and the temperature $T$. An energy equation, sometimes also called heat transport equation, is required to have a closed system. Here we adopt the approach of \citet{low1975} in which the energy equation is not solved directly. The key point is to select a temperature profile based on some observational constraints, in particular, we use the fact that in CHs the plasma temperature is lower than in the coronal environment; see \citet{munroetal1972} and the references in \citet{cranmer2009}. When we have obtained a solution, we calculate the corresponding energy balance that the system has to satisfy in order to keep a thermal equilibrium. Because it is difficult to solve the energy equation simultaneously with force balance, another option is to prescribe a priori a polytropic equation of state, which effectively corresponds to some energy addition or sink computed a posteriori (e.g. \citealt{tsinganosetal1992} and \citealt{petrieetal2002}).  Nevertheless, here we prefer to select an a priori temperature profile based on some observational constraints.

%, but this is not the main goal of the present study.
 
To avoid unnecessary complications, we restrict our analysis to a 2D coordinate geometry with invariance in the $y-$direction and with no component of the magnetic field in this direction. The magnetic field is written in terms of a magnetic flux function (the $y-$component of the magnetic vector potential), denoted by $A(x,z)$ and ensuring Eq.~(\ref{eq:divb}), but this needs to be determined.  We write
\begin{align}
    B_x(x,z)&=-\frac{\partial A}{\partial z}(x,z),\label{eqbxf}\\
    B_y(x,z)&=0,\label{eqbyf}\\
    B_z(x,z)&=\frac{\partial A}{\partial x}(x,z). \label{eqbzf}
\end{align}
The dependence of gas pressure and temperature on the magnetic flux function is prescribed according to the type of solution that we seek. It can be shown, see for example \citet{low1975}, \citet{priest1982} and the multiple examples explored in \citet{priestforbes2007}, that the flux function is the solution of the following non-linear elliptical partial differential equation:
\begin{eqnarray}
 \frac{\partial^2 A}{{\partial x^2}}(x,z)+\frac{\partial^2 A}{{\partial z^2}}(x,z)+\mu_0 \frac{\partial p(A,z)}{\partial A}=0, \label{eqgradshastatic}
\end{eqnarray}
where gas pressure is given by
\begin{align}
p(A,z)=p_0(A)\,e^{\textstyle -\int_0^z \frac{\bar{\mu} g }{\mathcal{R} T(A,z')}\,dz'},\label{eqpressalongB}
\end{align}
and $T(A,z)$ is the temperature profile that can depend on the flux function $A$ and the $z$ coordinate as well. Equation~(\ref{eqpressalongB}) imposes a balance between the force due to the gas pressure gradient and the gravity force along the magnetic field lines, while  Eq.~(\ref{eqgradshastatic}) represents the condition of force balance perpendicular to the magnetic field. The function $p_0(A)$ determines the profile of gas pressure at $z=0$. Equation~(\ref{eqgradshastatic}) is of Grad-Shafranov type, but includes the effect of gravity. This equation must be solved under some boundary conditions.

It is useful to rewrite Eq.~(\ref{eqgradshastatic}) in dimensionless form because it reduces to 
\begin{eqnarray}
 \frac{\partial^2 \bar A}{{\partial \bar x^2}}(\bar x,\bar z) +\frac{\partial^2 \bar A}{{\partial \bar z^2}}(\bar x,\bar z)+\frac{\beta_0}{2} \frac{\partial \bar p(\bar A,\bar z)}{\partial \bar A}=0, \label{eqgradshastaticadim}
\end{eqnarray}
where
\begin{align}
    \beta_0=\frac{2\mu_0 p_{00}}{B_0^2}
\end{align}
is the reference plasma-$\beta$, the lengths are normalised to $h$, the pressure to $p_{00}$ , and the flux function to $B_0 h$. The pressure scale height is $h=\mathcal{R} T_{\rm C}/ \bar{\mu} g$, where $T_{\rm C}$ is the reference coronal temperature ($h$ is typically about 60\,Mm for $T_{\rm C}=1$ MK). If gas pressure is ignored in Eq.~(\ref{eqgradshastaticadim}) ($\beta_0=0$), this equation reduces to a Laplace equation that leads to the potential solution. In this case, a maximum principle exists, stating that a solution cannot attain a maximum (or minimum) at an interior point of its domain. This result implies that the values of the solution in a bounded domain are bounded by its maximum and minimum values on the boundary.

The magnetic field and the thermodynamic variables are given in terms of $A$, $p_0(A),$ and $T(A,z)$. We distinguish between the photospheric
and coronal footpoint definition, which can be quite different in case of a canopy-like divergence from
the phothosphere to the base of the corona. Our reference level, $z=0$, is located at the base of the corona. If we know the magnetic field at this reference level, that is, $B_z(x,z=0)$, then we calculate the flux function at this level 
by direct integration of Eq.~(\ref{eqbzf}),
\begin{align}
A(x,z=0)=\int B_z(x,z=0)\, dx,\label{eq:Aintegral}
\end{align}
which is used to  calculate the gas pressure at the reference level
\begin{align}
p(x,z=0)=p_0(A(x,z=0)).
\end{align}
We need to solve Eq.~(\ref{eqgradshastatic}) subject to the boundary value of $A(x,z=0)$ (a Dirichlet problem) previously calculated, and in terms of some given $T(A,z)$ that has not been prescribed yet. The most relevant boundary condition is at the bottom of the system, $z=0$, this determines the behaviour of the solution in the whole spatial domain. The Grad-Shafranov equation is a fourth-order partial differential equation, meaning that we have to specify additional boundary conditions.  We have explored two situations, a finite rectangular domain and the upper half-plane (with the solution assumed to vanish at infinity), and decided to concentrate on the last configuration because it is not affected by the location of the edges of a finite domain. 

%This type of BC is relatively easy to apply analytically but more %difficult to implement from the numerical point of view. Details %about its numerical implementation are given later.

It is worth mentioning that \citet{low1980} proposed another scheme in which treating $A(x,z)$ as a known function then Eq.~(\ref{eqgradshastatic}) is used to determine $T(A,z)$. This scheme is more flexible and the equations of magnetostatic equilibrium can readily be integrated in closed form. However, for the purposes of the present paper related to the investigation of different profiles for $T(A,z)$, the scheme of \citet{low1975} adopted here is more convenient.

\section{Magnetic configuration}

In this section we calculate the magnetic field configuration. We begin with the simplest case, a unipolar magnetic field that represents a CH, and then we explore a bipolar model describing an AR. We provide rather simple analytical solutions in the case $\beta=0$, that is, the potential magnetic field.

\subsection{Unipolar CH and bipolar AR}\label{sect:CHARmodel}

\begin{figure}[h]
\center
\includegraphics[width=7.cm]{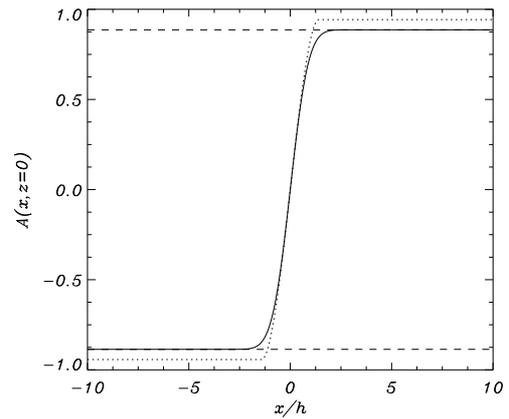}
\caption{\small Flux function as a function of the horizontal position at $z=0$. The continuous line corresponds to the Gaussian profile, and the dotted line shows the parabolic profile. The horizontal dashed lines represent the constant values that the Gaussian profile tends to asymptotically. In this plot, $w_0/h=1$, and $C=0$ for the Gaussian profile, while $x_0/h=1$ for the parabolic profile.} \label{figerfx}
\end{figure}

The type of solutions we are interested in should represent a CH, and the boundary conditions at $z=0$ allow us to chose families of solutions that have a vertical magnetic field at the center of the hole that progressively expands with distance from the central part. This is the sort of configuration that is inferred from the observations and from photospheric magnetic field extrapolations.  A convenient choice is a function that is concentrated at $x=0$ and that decreases with distance, for example following a Gaussian dependence,
\begin{eqnarray}
B_z(x,z=0)=B_0\,e^{-\left(\textstyle  \frac{x}{w_0}\right)^2}, \label{eqbz0}
\end{eqnarray}
where $w_0$ is the characteristic width, and the Gaussian is centered around $x=0$. Integrating with respect to $x$, Eq.~(\ref{eq:Aintegral}),  we find that the corresponding flux function is
\begin{eqnarray}
A(x,z=0)=\frac{\sqrt{\pi}}{2} B_0 w_0 \, {\rm Erf}\left(\frac{x}{w_0}\right)+C, \label{eqbz0partn}
\end{eqnarray}
where ${\rm Erf}(x)$ is the error function, and $C$ is an integration constant that is set to zero without loss of generality. The obtained flux function is proportional to the product $B_0\,w_0$ and to the error function, which is very similar to the hyperbolic tangent function. An example of the flux function is shown in Fig.~\ref{figerfx}. It is worth noting that although the magnetic field is localized in space, the flux function is not confined around $x=0$ and has tails that extend over the whole spatial domain. This is due to the fact that the magnetic field is unbalanced, that is, the net magnetic flux is not zero.

Other profiles for the magnetic field can be explored, but as we show in the following, the main features of the flux function are found to be essentially the same. We consider an inverse parabolic profile for the magnetic field,
\begin{eqnarray}\label{eq:parabbz}
B_z(x,z=0)=\begin{cases}
B_0-B_0 \left(\textstyle \frac{x}{x_0}\right)^2&-x_0<x<x_0,\\
0&|x|\geq x_0.
\end{cases}
\end{eqnarray}
The corresponding flux function is 
\begin{eqnarray}\label{eq:aparabf}
A(x,z=0)=B_0 F(x)=\begin{cases}
B_0 x -\frac{B_0}{x_0^2} \frac{x^3}{3}&-x_0<x<x_0,\\
\frac{2}{3}B_0 x_0  &x\geq x_0
,\\
-\frac{2}{3}B_0 x_0 &x\leq -x_0,
\end{cases}
\end{eqnarray}
where the values of $A(x,z=0)$ for $|x|$ higher than or equal to $x_0$ (we assume that $x_0>0$) were adjusted through the integration constant to have a continuous derivative of $A$ and therefore a continuous variation of $B_z(x,z=0)$. The integration constant $C$ has also been set to zero. We introduced the definition of $F(x)$ in Eq.~(\ref{eq:aparabf}). The flux function is plotted in Fig.~\ref{figerfx} and  can be compared with the corresponding flux function associated with the Gaussian profile. The two profiles are very similar, meaning that for our purposes, the specific function we chose to represent the unipolar magnetic field is unimportant. In the rest of this section, we use the parabolic profile for the magnetic field and  the corresponding flux function. The reason for choosing this dependence is that, as we show below, analytical progress is possible if the polynomial form of the flux function given by Eq.~(\ref{eq:aparabf}) is used. When the energetics of the problem is addressed in Sect.~\ref{sect:energy}, the Gaussian profile is more convenient because the conduction term (involving second-order spatial derivatives) has a smoother behaviour than the term for the parabolic profile (which has discontinuous derivatives in the $x-$direction).  

%\subsection{The bipolar configuration: active region}

\begin{figure}[h]
\center
\includegraphics[width=7.cm]{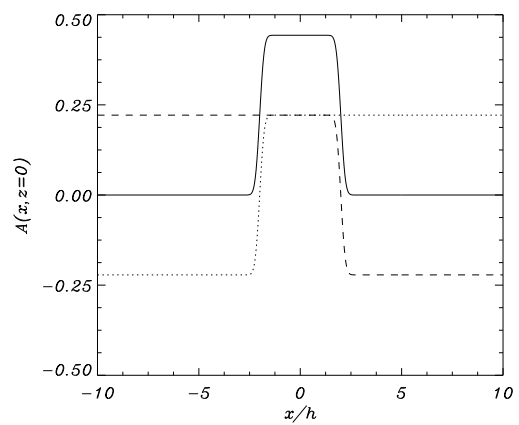}
\caption{\small Flux function as a function of the horizontal position at $z=0$ for the bipolar AR described by Eq.~(\ref{eqApartAR}). The continuous line corresponds to the superposition of the left unipolar region (dotted line) and the right unipolar region (dashed line). In this plot, $x_1/h=-2$, $x_2/h=2$ and $x_0/h=1/4$.} \label{figerfxbipol}
\end{figure}

Now we construct an elementary magnetic model that represents an AR. Such configurations are in general bipolar, and a straightforward method to represent them is to superpose two individual unipolar regions of opposite polarity and separated by a certain distance. In terms of the function $F(x)$ introduced before, the flux function is
\begin{align}
A(x,z=0)=B_1 F(x-x_1)+B_2 F(x-x_2),
\label{eqApartAR}
\end{align}
where we assume hereafter that $B_1 B_2<0$ to have a bipolar configuration. The parameters $x_1$ and $x_2$ correspond to the centres of the opposite-polarity unipolar magnetic fields. The distance between them therefore is $|x_1-x_2|$. If we assume that the two unipolar regions are equal (same widths and intensities, except  for the sign of the magnetic field), then the flux function, in contrast to the case of a unipolar field, is a localized function in space. In this situation, there is a perfect balance that cancels the tails of the flux function. This feature is clear in Fig.~\ref{figerfxbipol}.

In a general case, it is easy to see that the minimum and maximum values of the flux function are given by 
\begin{align}
A_{min}=\left(-B_1 x_0-B_2 x_0\right)\frac{1}{2},
\end{align}
while the absolute maximum value is
\begin{align}
A_{max}=\left(B_1 x_0-B_2 x_0\right)\frac{1}{2}.
\end{align}
We assumed that the thickness of each unipolar region is the same, $x_0$, but the previous expressions are easily extended to non-equal widths. The minimum and maximum values of the flux function are useful when the thermal structure of the model is introduced in Sect.~\ref{sec:modelI}.

\subsection{Solution for the upper half-plane: Green's function}\label{sect:green}

 For the situation $\beta=0,$ we showed  that the Grad-Shafranov equation given by Eq.~(\ref{eqgradshastatic}) reduces to a Laplace equation. We assume that the domain is $z>0$ (upper half-plane) and therefore we implicitly assume that the flux function tends to zero for $x\rightarrow \pm \infty$ and $z\rightarrow \infty$. For this kind of problem, it is known that we can use the properties of Green's functions.  They have been applied in the past to calculate potential magnetic fields based on photospheric magnetograms (see \citealt{schmidt1964} and \citealt{sakurai1982} for details about the first applications of the Green function method in the solar context). When this method is used, it is known that the Laplace equation in the Cartesian domain $z>0$ plus the inhomogeneous boundary condition at $z=0$ have the following analytical solution:
\begin{align}\label{eq:green0}
A_0(x,z)=\frac{z}{\pi}\int_{-\infty}^{\infty}\frac{A(\xi,z=0)}{(x-\xi)^2+z^2}\, d\xi, \end{align}
where we use the subindex 0 to indicate that it corresponds to the zero $\beta$ case. 
This exact solution tends to zero at large distances from the source, and it involves a definite integral and the specific profile of the flux function at $z=0$, that is, $A(x,z=0)$. In general, the integral in Eq.~(\ref{eq:green0}) needs to be calculated numerically, but for our deliberate choice of the parabolic profile for the unipolar field at $z=0$ given by Eq.~(\ref{eq:aparabf}), the primitive is analytical,
\begin{align} \label{eq:a0unip}
A_0(x,z)&=\frac{a_0}{4 \pi x_0^3}\left(z \left(-3 x^2+3 x_0^2+z^2\right) \ln \frac{(x-x_0)^2+z^2}{(x+x_0)^2+z^2}\right.\nonumber \\
&+\left(2 x^3-6 x x_0^2-6 x z^2+4 x_0^3\right) \tan ^{-1}\left(\frac{x-x_0}{z}\right)\nonumber \\
&\left. +\left(-2 x^3+6 x x_0^2+6 x z^2+4 x_0^3\right) \tan ^{-1}\left(\frac{x+x_0}{z}\right)-8 x x_0 z\right),
\end{align}
where $a_0=(2/3)x_0 B_0$. This solution contains up to third-order polynomials together with logarithmic and trigonometric functions, and it is therefore relatively easy to handle. It can be shown that applying l'H\^opital's rule, the limit of Eq.~(\ref{eq:a0unip}) when $z$ tends to $\infty$ is zero, as expected.

The components of the magnetic field associated with Eq.~(\ref{eq:a0unip}) using Eqs.~(\ref{eqbxf}) and (\ref{eqbzf}) are\begin{align} \label{eq:bxunip}
B_x(x,z)=\frac{3 a_0}{4 \pi x_0^3}\Bigg(4 x x_0  &+\left(x^2-x_0^2-z^2\right)  \ln \frac{(x-x_0)^2+z^2}{(x+x_0)^2+z^2} \nonumber \\
 &+4x z\left[ \tan ^{-1}\left(\frac{x-x_0}{z}\right)- \tan ^{-1}\left(\frac{x+x_0}{z}\right)\right]\Bigg),
\end{align}
\begin{align} \label{eq:bzunip}
B_z(x,z)=&-\frac{3 a_0}{2 \pi x_0^3}\Bigg(2 x_0 z + x z \ln \frac{(x-x_0)^2+z^2}{(x+x_0)^2+z^2} \nonumber \\
 &+\left(-x^2+x_0^2+z^2\right)\left[ \tan ^{-1}\left(\frac{x-x_0}{z}\right)- \tan ^{-1}\left(\frac{x+x_0}{z}\right)\right]\Bigg).
\end{align}
Substituting $z=0$ in Eq.~(\ref{eq:bzunip}) and using that $\tan^{-1} (\pm \infty)=\pm \pi/2,$ we find that only the third term on the right-hand side is different from zero, and we recover the parabolic profile given by Eq.~(\ref{eq:parabbz}) that was used as boundary condition to construct the solution in the whole 2D domain.

\begin{figure}[h]
\center
{\includegraphics[width=7.cm]{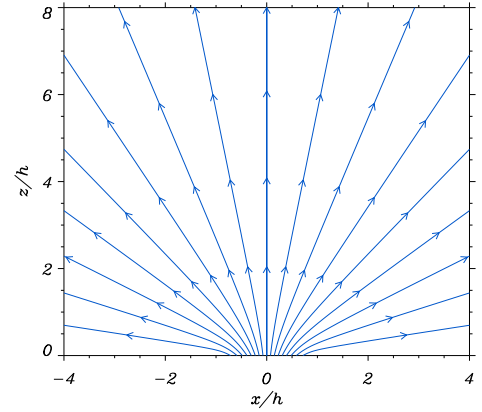}
}
\caption{\small Magnetic field lines in the analytical CH model given by Eq.~(\ref{eq:a0unip}) ($\beta_0=0$). In this example, the half-width of the CH at $z=0$ is $x_0=h$. Magnetic field lines coincide with isocontours of the flux function $A_0$. The geometry of the magnetic field would be affected by boundaries at a finite distance from the source, but the method implemented in this work prevents these effects through the Green functions. } \label{figCHbeta0}
\end{figure}

An example of a CH model  based on the previous analytical solution is shown in Fig.~\ref{figCHbeta0} for a particular choice of the parameters $B_0$ and $x_0$. The magnetic field is purely vertical at $x=0$ and globally shows a  radial expansion of straight magnetic field lines except near the bottom boundary, where the field matches the boundary conditions at $z=0$ and the magnetic field lines are slightly curved.  At low heights and far from the source, the horizontal component of the magnetic field dominates the vertical component. Because at $z=0$ we impose that $B_z(x,0)=0$ for $|x|\ge x_0$ (see Eq.~(\ref{eq:parabbz})), the magnetic field becomes purely horizontal at our reference level. In this regard, the Gaussian profile given by Eq.~(\ref{eqbz0}) would avoid this effect and produce a more confined structure. In a real CH, it is quite unlikely that the magnetic field is purely horizontal outside the CH because there are contributions from other nearby magnetic fields with different spatial scales and intensities that have emerged through the photosphere (this is partially addressed in the following paragraphs, where a CH is combined with an AR).

\begin{figure}[h]
\center
{\includegraphics[width=7.cm]{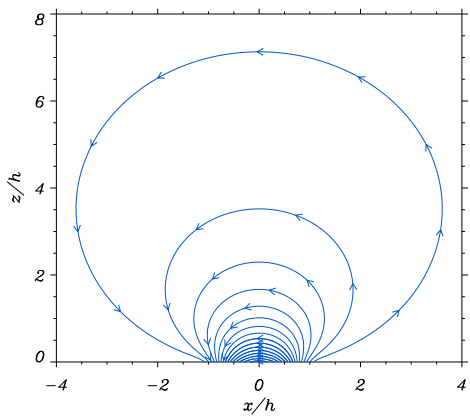}
}
\caption{\small Magnetic field lines in a symmetric AR model based on Eq.~(\ref{eqApartAR}). In this example, the two unipolar sources are located at $x_1=-1/4h$ and $x_2=1/4h,$  $\text{where }x_0=h$.} \label{figARbeta0}
\end{figure}

\begin{figure}[h]
\center
{\includegraphics[width=7.cm]{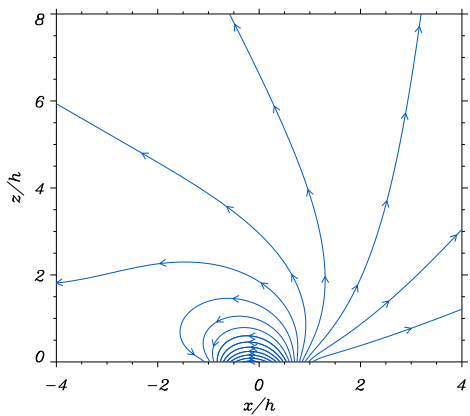}
}
\caption{\small Magnetic field lines in a non-symmetric AR model. In this example, the two unipolar sources are located at $x_1=1/4h$ and $x_2=-1/4h,$ where $x_0=h$. The source located at $x=x_1$ is 1.2 times stronger than the source at $x=x_2$. There is some spatial overlap in the sources, but this is allowed in our model because it is based on the superposition of individual solutions.} \label{figARnosymbeta0}
\end{figure}

The analytical solution for the unipolar magnetic field is the basis for constructing more complicated magnetic configurations. As we described in the previous section, we build an AR model by adding two unipolar magnetic fields of opposite polarity.  We can use this superposition of solutions for the flux function because the Laplace equation is a linear partial differential equation. We  introduce a translation of the solution given by Eq.~(\ref{eq:a0unip}) with respect to the $x-$coordinate by making the change $x\rightarrow (x-x_1)$, where $x_1$ is the centre of the new unipolar region. We repeated this operation for the opposite-polarity unipolar field, located at $x_2$ , and then we added the corresponding flux functions according to Eq.~(\ref{eqApartAR}). An example of this equilibrium is shown in Fig.~\ref{figARbeta0}. We obtain closed field lines that for our purposes represent a typical magnetic configuration of a symmetric bipolar AR. This magnetic configuration was used later to host a dense and hot plasma in the case of a plasma-$\beta$ different from zero. Figure \ref{figARnosymbeta0} shows a similar situation, but the AR is not perfectly symmetric: the magnetic field at the right footpoint is stronger than that at the left footpoint. At low heights, the configuration is closed, but it changes to open away from the AR core. At very large distances from the bipolar region, we essentially recover the structure of a single unipolar configuration with straight magnetic field lines; compare with Fig.~\ref{figCHbeta0}.

\begin{figure}[h]
\center
{\includegraphics[width=7.cm]{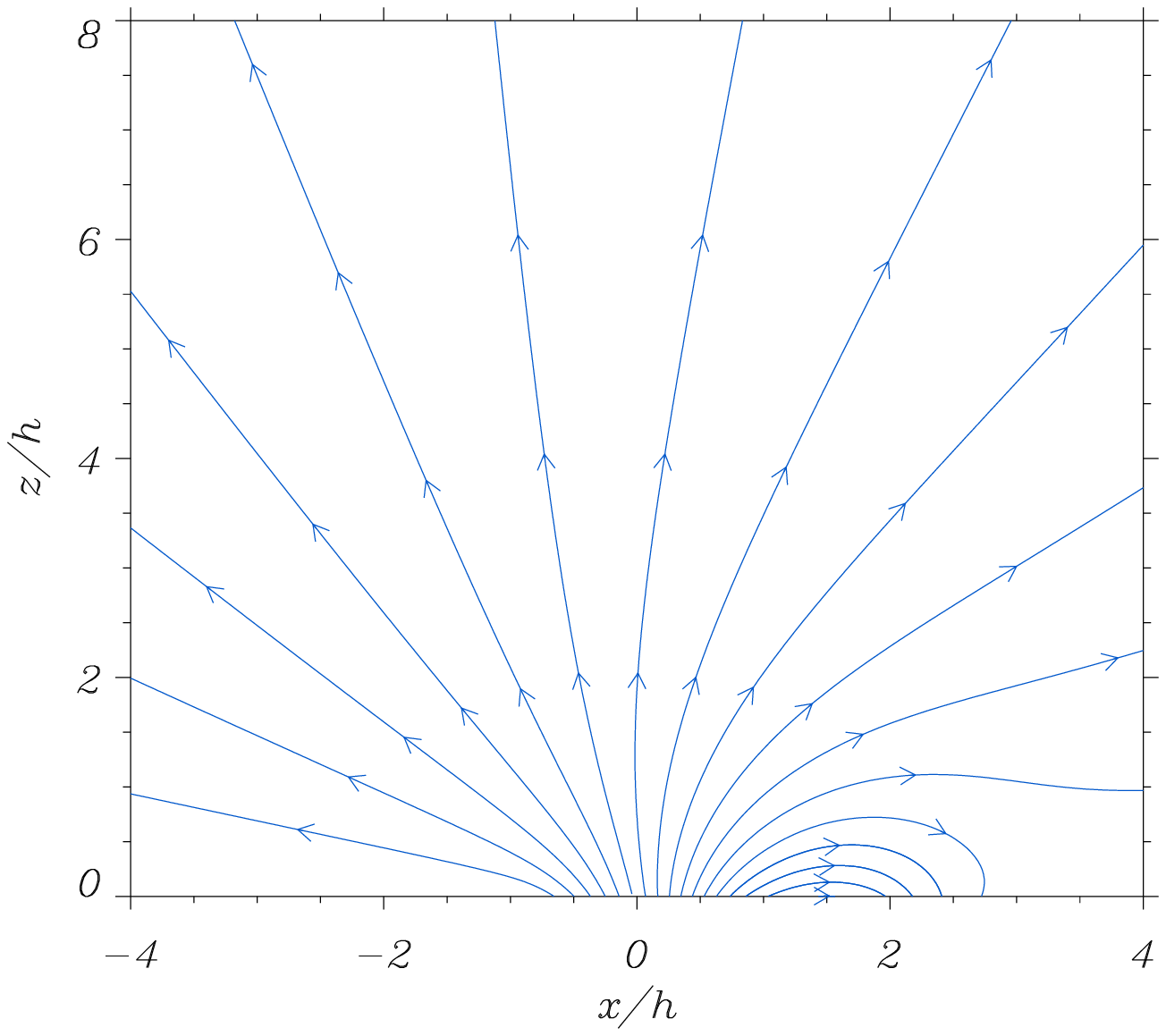}
\includegraphics[width=7.cm]{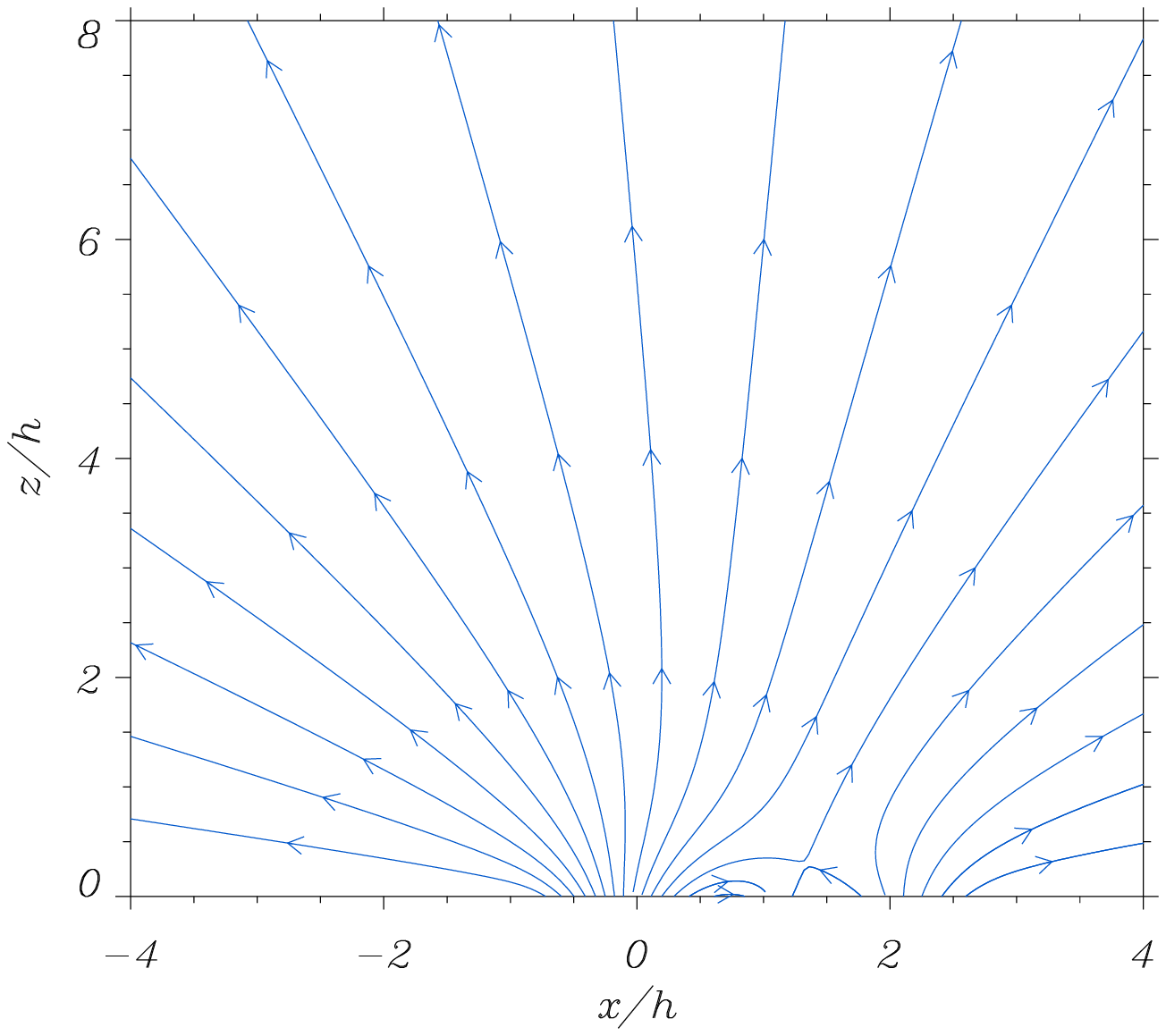}
}
\caption{\small Magnetic field lines of a mixed CH and  AR model. The CH is located at $x_1=0,$ and the AR is represented by equal opposite-polarity sources located at $x_1=h$ and $x_2=2h,$ where $x_0=h$. The differences between the two panels are due to the polarity of the left footpoint of the AR with respect to the CH: positive in the left panel, and negative in the right panel. This creates an X-point. } \label{figAR+CH}
\end{figure}

Two additional examples of equilibrium based on  the superposition of unipolar regions are shown in Fig.~\ref{figAR+CH}. The magnetic configuration combines a CH and an AR. In the first case, the polarity of the left footpoint of the AR is the same as the polarity of the CH. The CH is displaced towards the left due to the presence of the AR, which is squashed by the CH. Due to the presence of the AR, the magnetic field is no longer purely horizontal at the right part of the CH for $x>x_0,$ as happens in the configuration of Fig.~\ref{figCHbeta0}. In the second situation, the polarity of the AR is reversed. The CH is shifted to the right, and a null or X-point, with zero magnetic field, appears in the AR. In this configuration, the separatrixes of the magnetic field provide an excellent definition of the location of the CHB (understood as the layer that separates open from closed magnetic field lines). In conclusion, the examples shown in this section illustrate the flexibility of the scheme to construct a variety of  potential magnetic field configurations that can be used in general as the basis for other studies, and in particular in the following section as the skeleton to include the effect of gas pressure and gravity in the system.

\section{Coupling the magnetic field to the plasma }\label{sec:modelI}

When the $\beta$ of the plasma is different from zero, the magnetic field and the plasma are coupled. Because this parameter in CHs, ARs, and in general in the corona is small, the deviation from the potential solution is expected to be limited. Nevertheless, it is crucial to include gas pressure in the scheme because the observations specifically provide information about the structure in density (emission measure) and temperature (both related to gas pressure), and to a lesser extent, about the magnetic field, inferred from extrapolations based on photospheric  magnetograms. 

%The information about the heating is only deduced %indirectly from the observations when a certain %theoretical model is assumed. 

We included a non-zero gas pressure in Eq.~(\ref{eqgradshastatic}) through the in principle arbitrary functions $p_0(A)$ and $T(A,z)$ that need to be defined. For simplicity and with the aim of focusing on the overall behaviour, we assumed the following separable dependence for temperature, although this is not mandatory in the present formalism:
\begin{align}
T(A,z)=\mathcal{T}(A)\mathcal{H}(z).\label{eq:Tsep}
\end{align}
This profile entails a type of uniform temperature dependence on height in the whole region of interest, either the CH or AR, but temperature is still allowed to vary from field line to field line due to the term $\mathcal{T}(A)$. This situation is similar to the cases of self-similar MHD solutions that were explored, for example, by \citet{petrieetal2002,petrieetal2003} and \citet{tsinganos2010}. An isothermal temperature distribution along a given field line ($A=\rm const$) is obtained by imposing that $\mathcal{H}(z)=1$. In this case, and according to Eq.~(\ref{eqpressalongB}), the gas pressure dependence is
\begin{align}
p(A,z)=p_0(A)\,e^{\textstyle -\frac{\bar{\mu} g }{\mathcal{R} {\mathcal T}(A)} z},\label{eqpressalongBconst}
\end{align}
which is the well-known exponential stratification whose scale height depends on the temperature of the plasma along the field line that we consider (the particular value of $A$).

For the non-isothermal situation, a simple choice for the temperature dependence with height is
\begin{align}\label{eq:h}
\mathcal{H}(z)=e^{z/\Lambda},
\end{align}
where $\Lambda$ is the characteristic spatial scale, which might be positive or negative. This allows us to consider a situation with an increasing or decreasing temperature with height and therefore with a net effect of  thermal conduction that changes the energy balance in the system; this is addressed in Sect.~\ref{sect:energy}. The pressure dependence in the non-isothermal case calculated from Eq.~(\ref{eqpressalongB}) is 
\begin{align}
p(A,z)=p_0(A)\,e^{\textstyle \frac{\bar{\mu} g }{\mathcal{R} \mathcal {T}(A)} \Lambda\, (e^{-z/\Lambda}-1)},\label{eqpressalongBnoniso}
\end{align}
and deviates from the easy exponential dependence with $z$ for the isothermal situation, Eq.~(\ref{eqpressalongBconst}).

According to these equations, the gas pressure term  that appears in Eq.~(\ref{eqgradshastatic}) for the isothermal case (when we refer to an isothermal situation, we mean that temperature does not change along the magnetic field lines, but it changes from line to line) reads \begin{eqnarray}
\frac{\partial p}{\partial A}(A,z)=e^{\textstyle -z\frac{\bar{\mu} g}{\mathcal{R} {\mathcal T}(A) }}\left(\frac{\partial p_0}{\partial A}(A)+z \frac{\bar{\mu} g}{\mathcal{R}}\frac{\partial \mathcal T}{\partial A}(A)\frac{p_0(A)}{{\mathcal T}^2(A)}\right),\label{eqderivP}
\end{eqnarray}
while for the non-isothermal case, using Eq.~(\ref{eqpressalongBnoniso}), we simply use the transformation
\begin{eqnarray}\label{eqderivPnoiso}
z\rightarrow -\Lambda\, (e^{-z/\Lambda}-1),
\end{eqnarray}
because the derivatives in Eq.~(\ref{eqderivP}) do not affect the $z$ coordinate explicitly.

\subsection{CH and AR thermal structure}

When some general features of the pressure and temperature dependence in our model are known, we must constrain the functional form of $p_0(A)$ and ${\mathcal T}(A)$ according to observations of CH and AR. For the CH model, we chose these functions to have a minimum at the centre of the structure and to increase smoothly with position to match coronal values,  creating a depletion in pressure and temperature inside the CH. This is the common behaviour inferred from observations of CHs \citep[see e.g.][]{cranmer2009}. Our specific choice was
\begin{align}
p_0(A)=\left(p_{\rm C}-p_{\rm CH}\right)\left(\frac{A}{A_{ref}}\right)^2+p_{\rm CH},\nonumber \\
{\mathcal T}(A)=\left(T_{\rm C}-T_{\rm CH}\right)\left(\frac{A}{A_{ref}}\right)^2+T_{\rm CH}, \label{eqp0t}
\end{align}
where $p_{\rm C}$ and $T_{\rm C}$ are the reference coronal pressure and temperature values outside the CH, and $p_{\rm CH}$ and $T_{\rm CH}$ are the values inside the CH (satisfying that $p_{\rm CH}/p_{\rm C}< 1$ and $T_{\rm CH}/T_{\rm C}<1$ to have representative CH conditions). In Eq.~(\ref{eqp0t}) the dependence on the square of the flux function, $A$, arises because firstly, pressure and temperature must be positive defined (and $A$ is not necessarily positive everywhere, see Fig.~\ref{figerfx}), and secondly, the unipolar region should be symmetric with respect to the centre of the CH because the magnetic field is symmetric (the parabolic profile). The quadratic dependence satisfies these requirements, but this is not a uniquely possible choice. In 
Eq.~(\ref{eqp0t}) the flux function $A$ is divided by a reference value, $A_{ref}$, which in this present case is chosen to be equal to $A_{min}$, previously introduced in Sect.~\ref{sect:CHARmodel}. This allows us to fix coronal values to $p_{\rm C}$ and $T_{\rm C}$ for $A=A_{min}$. The values of $p_{\rm CH}$ and $T_{\rm CH}$ are achieved when $A=0$, that is, when the magnetic field is vertical in our configuration. Applying the ideal gas law, we obtain the relation 
\begin{align}
\frac{\rho_{\rm CH}}{\rho_{\rm C}}=\frac{p_{\rm CH}}{p_{\rm C}} \frac{T_{\rm C}}{T_{\rm CH}},
\end{align}
which must be smaller than one to represent a CH (under-dense structure with respect to the environment). This means that we have the restriction $p_{\rm CH}/p_{\rm C}< T_{\rm CH}/T_{\rm C}$ when we choose the parameters of the CH model, that is, the pressure decrement must be smaller that the temperature decrement.

Other profiles in Eq.~(\ref{eqp0t}) can be adopted, but the important property is that the central values must be lower than the coronal values to properly represent CH conditions. It is necessary to remark that if we assume a constant temperature along the field lines, then thermal conduction, in principle relevant in the solar corona, would not have any effect because it is proportional to the second derivative of temperature with space, which is zero in this particular case.

The question that arises now is how we can define the dependence of pressure and temperature when an AR is considered. ARs are denser and hotter than the surrounding corona. According to the dependence of the flux function on position in a bipolar AR, see Fig.~\ref{figerfxbipol}, a suitable choice is
\begin{align}
p_0(A)=\left(p_{\rm AR}-p_{\rm C}\right)\left(\frac{A}{A_{ref}}\right)^2+p_{\rm C},\nonumber \\
{\mathcal T}(A)=\left(T_{\rm AR}-T_{\rm C}\right)\left(\frac{A}{A_{ref}}\right)^2+T_{\rm C}. \label{eq:arprof}
\end{align}
We could select a linear dependence with $A$ instead, but if the bipolar region has a small imbalance in the parameters $B_1$ and $B_2$ , then the flux function can be negative, and this is not convenient to ensure positive defined pressures and temperatures. With a quadratic dependence, we avoid these issues. Using 
Eq.~(\ref{eq:arprof}) when $A=0$ pressure and temperature have exactly coronal values, $p_{\rm C}$ and $T_{\rm C}$. At the centre of the AR, we have that $A=A_{ref}$ , and therefore pressure and temperature tend to the core values, $p_{\rm AR}$ and $T_{\rm AR}$ ($p_{\rm AR}/p_{\rm C}> 1$ and $T_{\rm AR}/T_{\rm C}>1$ to describe an AR). For a symmetric bipolar region, we have that $A_{ref}=A_{max}$ and $A_{min}=0$. Again, from the ideal gas law, we find that
\begin{align}
\frac{\rho_{\rm AR}}{\rho_{\rm C}}=\frac{p_{\rm AR}}{p_{\rm C}} \frac{T_{\rm C}}{T_{\rm AR}},
\end{align}
which must be greater than one to represent an over-dense region with respect to the environment. Therefore, the condition $p_{\rm AR}/p_{\rm C}> T_{\rm AR}/T_{\rm C}$  needs to be satisfied.

From Eqs.~(\ref{eqp0t}) and (\ref{eq:arprof}), we have some freedom to choose the main parameters of the models. Namely, $p_{\rm CH}$ and $T_{\rm CH}$ (satisfying $p_{\rm CH}/p_{\rm C}< T_{\rm CH}/T_{\rm C}$) for the CH model and $p_{\rm AR}$ and $T_{\rm AR}$ (satisfying $p_{\rm AR}/p_{\rm C}> T_{\rm AR}/T_{\rm C}$) for the AR model. A modification of these parameters affects the density values and eventually the magnetic field structure due to the coupling between the plasma and the magnetic field. We have an infinite number of possible solutions. 

It is indispensable to remark that the assumed functional dependencies for pressure and temperature for both the CH and the AR are essentially the same.  Equations~(\ref{eqp0t}) and (\ref{eq:arprof}) have the same dependence on the flux function $A$ (but $A$ is dissimilar for each structure), and the  differences are caused by the constant factors. For the CH model, however, the multiplicative factor is always positive ($p_{\rm C}-p_{\rm CH}>0$), and the same applies to the AR model ($p_{\rm AR}-p_{\rm C}>0$). This suggests that although the two structures may seem to have very dissimilar physical properties, they can be described using a common picture. Thus, the nature of CH and AR is not so antagonistic, according to our uninvolved representation. In this regard, a magnetic configuration containing both a CH and an AR might be composed, like those shown in Fig.~\ref{figAR+CH}, and the common thermal structure described by Eqs.~(\ref{eqp0t}) or (\ref{eq:arprof}) would produce depletions or enhancements in the density and temperature of the CH and AR contained in the system. This approach can be applied, for example, to investigate the coupling between CHs and ARs, but this topic is beyond the scope of the present work.

\subsection{Approximate semi-analytical method for computing equilibria in the low$-\beta$ regime}\label{sec:semi-an}

When we have chosen the specific profiles for $p_0(A)$ and $T(A,z)$ the Grad-Shafranov equation needs to be solved to understand how the topology of the magnetic field changes according to the values of the plasma parameters.  In general, finding solutions to Eq.~(\ref{eqgradshastatic}) requires numerical methods unless the profile for $p(A,z)$ and the boundary conditions are simple.  It is known that only when Eq.~(\ref{eqgradshastatic}) is a linear equation, are general analytical methods available. In the past, significant effort has been devoted to find exact analytical solutions in similar problems because they provide deep insights into the physics of the problem (see e.g. the extensive literature about the subject of B. C. Low). Before solving Eq.~(\ref{eqgradshastatic}) by purely numerical means in Sect.~\ref{sec:numres}, we introduce a semi-analytical method based precisely on the linearisation of this equation. First, the non-linear elliptic equation for the flux function is rewritten as
\begin{eqnarray}
 \frac{\partial^2 A}{{\partial x^2}}+\frac{\partial^2 A}{{\partial z^2}}+\epsilon f(A)=0. \label{eq:gradepsilon}
\end{eqnarray}
The parameter $\epsilon$ can be viewed as the plasma-$\beta$ (see the equivalent  non-dimensional Eq.~(\ref{eqgradshastaticadim})), and $f(A)$ is the derivative of the pressure term with respect to $A$, given by Eq.~(\ref{eqderivP}) for the isothermal case, or Eq.~(\ref{eqderivP}) modified according to  Eq.~(\ref{eqderivPnoiso}) for the non-isothermal case.

To make analytical progress, we use a pertubational expansion and write
\begin{align}\label{eq:Asum}
A=A_0+\epsilon A_1+\epsilon^2 A_2+ ...
\end{align}
We obtain the following equation to zero order in $\epsilon:$
\begin{eqnarray}
 \frac{\partial^2 A_0}{{\partial x^2}}+\frac{\partial^2 A_0}{{\partial z^2}}=0, \label{eq:gradepsilona0}
\end{eqnarray}
which is strictly the Laplace equation for $A_0$ that has been solved analytically in Sect.~\ref{sect:green} given a superposition of parabolic magnetic field profiles at $z=0$. The solution to this equation is just the potential magnetic field.

Now we apply a Taylor expansion to $f(A),$ keeping terms up to first order in $\epsilon$,
\begin{align}
f(A)\approx f(A_0+\epsilon A_1)\approx f(A_0)+\epsilon A_1 f'(A_0).
\end{align}
Using this expansion in Eq.~(\ref{eq:gradepsilon}), we find, to first order in $\epsilon$, the following equation:
\begin{eqnarray}
 \frac{\partial^2 A_1}{{\partial x^2}}+\frac{\partial^2 A_1}{{\partial z^2}}=-f(A_0). \label{eq:gradepsilona1}
\end{eqnarray}
 Equation~(\ref{eq:gradepsilona1}) is just a Poisson equation for $A_1$ and the inhomogeneous or source term on the right-hand side depends on the solution $A_0$ to the Laplace equation. Appropriate boundary conditions need to be applied, but the inhomogeneous BC at $z=0$ has been incorporated in the Laplace solution, and therefore, homogeneous BCs (i.e. $A_1=0$) need to be imposed on the Poisson solution so that the full solution given by Eq.~(\ref{eq:Asum}) satisfies the required BC at $z=0$. Hence, the highly non-linear problem has so far been reduced to solving a linear Laplace equation plus a Poisson equation, which is also linear. We can keep higher-order terms in $\epsilon$, and to second order, we have
\begin{eqnarray}
 \frac{\partial^2 A_2}{{\partial x^2}}+\frac{\partial^2 A_2}{{\partial z^2}}=-A_1 f'(A_0), \label{eq:gradepsilona2}
\end{eqnarray}
obtaining again a Poisson equation, but the source term depends on $A_1$ and $A_0$, that is, the previous order solutions.  Nevertheless, in this work we only consider terms up to first order in $\epsilon$.

The method of images used to solve the Laplace equation in Sect.~\ref{sect:green} is also applied here to solve the Poisson equation given by Eq.~(\ref{eq:gradepsilona1}), but forcing the Green function to be zero at $z=0$ because of the required homogenous BC. In this situation, we have to use a source image of the point $(x,z)$ with respect to the $x-$axis, $(x,-z)$. The analytical solution to the Poisson equation in this case is known to be  \citep[e.g.][]{myintu2009linear}
\begin{align}\label{eq:green2D}
A_1(x,z)=-\frac{1}{4 \pi}\int_{0}^{\infty} \int_{-\infty}^{\infty} f(A_0(\xi,\eta)) \ln{\frac{(x-\xi)^2+(z-\eta)^2}{(x-\xi)^2+(z+\eta)^2}}\, d\xi\,d\eta. \end{align}
This formal solution to the Poisson equation involves the evaluation of a double integral, and in general, it is rather difficult to obtain a closed analytical form of the solution in terms of known functions.

A purely numerical evaluation of Eq.~(\ref{eq:green2D}) is required for the examples investigated in the present paper. For this reason, we provide the main steps to evaluate Eq.~(\ref{eq:green2D}) numerically without major difficulties. First, we use the following result for the inner integral: $\int_{-\infty}^{\infty} \mathcal{F}(y)\, dy=\int_{0}^{\infty} \left[\mathcal{F}(y)+\mathcal{F}(-y)\right] dy$. Then we introduce a variable transformation to have finite limits in the inner and outer integrals.  A convenient variable transformation is $\xi=(1-t)/t$ and $\eta=(1-s)/s,$ which changes the integration domain from $0<\xi<\infty$ and $0<\eta<\infty$ to the more convenient range $0<s<1$ and $0<t<1$. The double integral in terms of the new variables reads
\begin{align}\label{eq:A1double}
&A_1(x,z)=\nonumber\\ &-\frac{1}{4 \pi}\int_{0}^{1}\int_{0}^{1}\bigg[ f\left(A_0\left(\frac{1-t}{t},\frac{1-s}{s}\right)\right) \ln{\frac{\left(x-\frac{1-t}{t}\right)^2+\left(z-\frac{1-s}{s}\right)^2}{\left(x-\frac{1-t}{t}\right)^2+\left(z+\frac{1-s}{s}\right)^2}}
\nonumber \\ &+f\left(A_0\left(\frac{-1+t}{t},\frac{1-s}{s}\right)\right) \ln{\frac{\left(x-\frac{-1+t}{t}\right)^2+\left(z-\frac{1-s}{s}\right)^2}{\left(x-\frac{-1+t}{t}\right)^2+\left(z+\frac{1-s}{s}\right)^2}}\bigg]\,\frac{1}{t^2} dt \frac{1}{s^2}\,ds.\nonumber \\ \end{align}
An examination of this double integral reveals that the points $t=0$ and $s=0$ need special treatment in order to avoid (regular) singularities. This is accomplished by choosing a quadrature formula based on an open method that does not explicitly use the end points (located at 0 and 1 in our case, but 1 is not problematic) to calculate the integral. Trapezoidal and Simpson methods do not fall into this category, but the standard composite midpoint method of integration is suitable. We find that this last method, with typically $200\times 200$ points, is enough to provide accurate values of the integral and convergence of the results is warrantied.

\begin{figure}[!h]
\center
{\includegraphics[width=7cm]{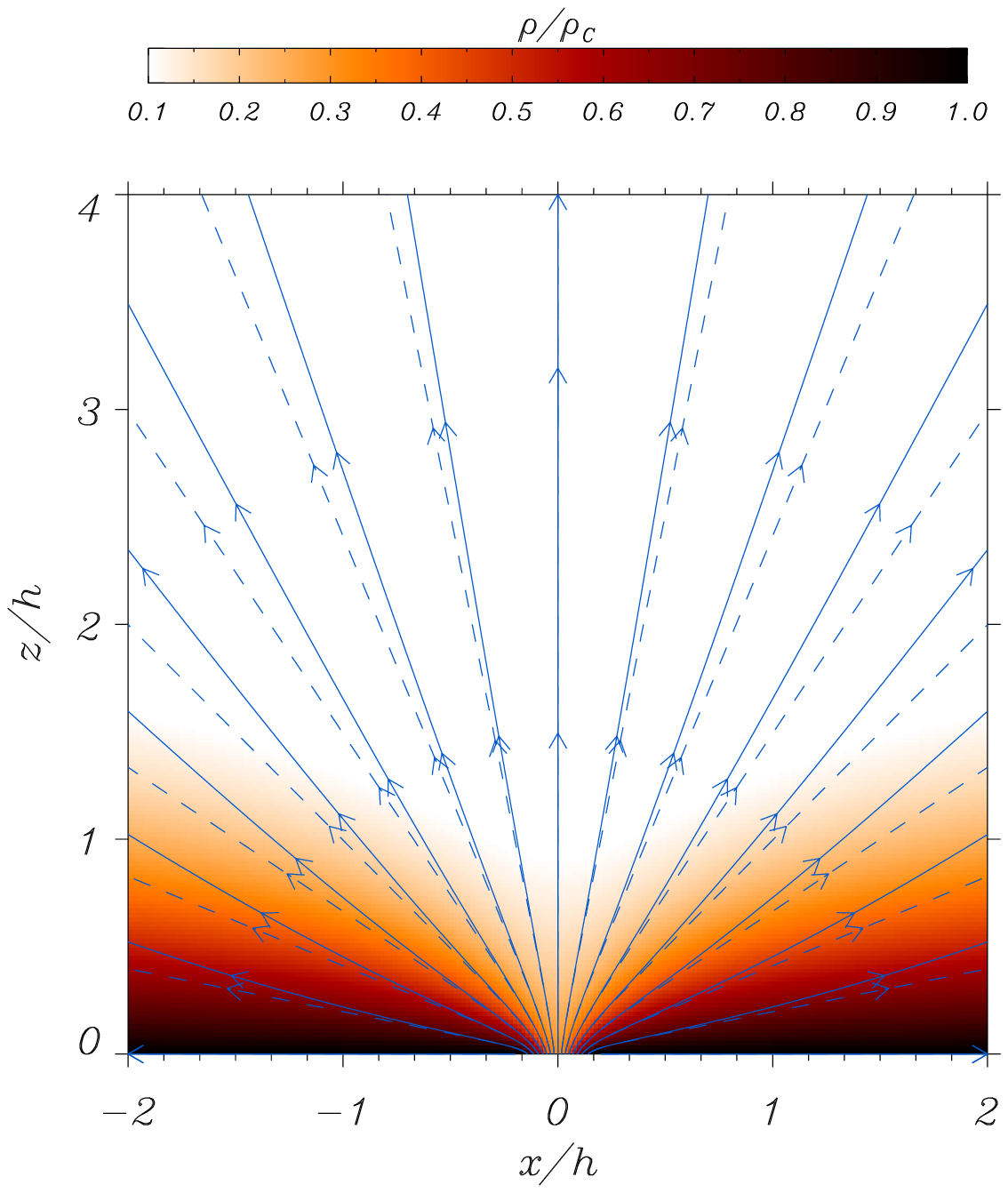}
\includegraphics[width=7cm]{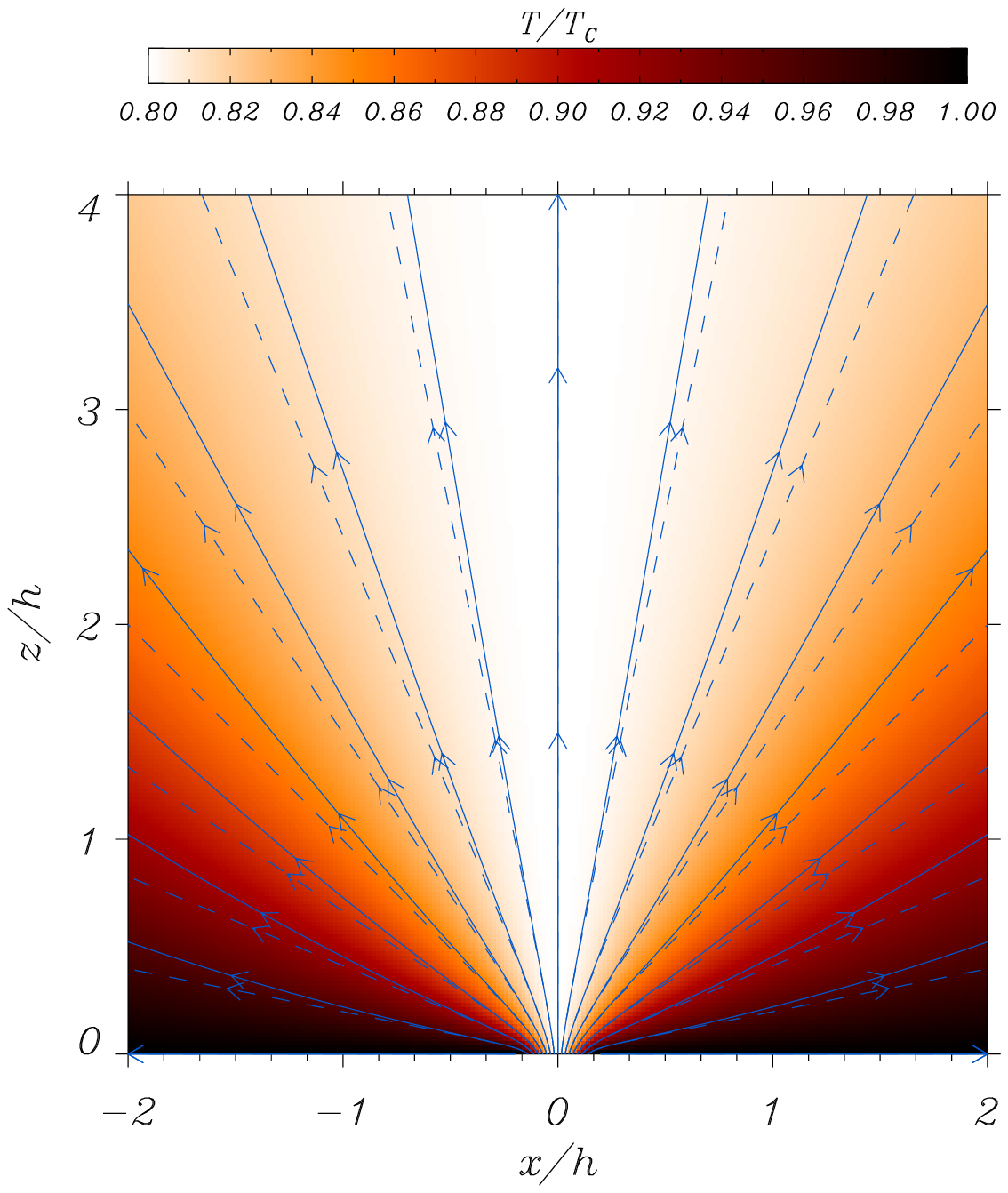}
}
\caption{\small Density and temperature distribution for the CH model with $\beta_0=0.004$. In this solution, $p_{\rm CH}/p_{\rm C}=1/4$, $T_{\rm CH}/T_{\rm C}=0.8$, $x_0/h=0.2$. The isothermal condition (along the field lines) is imposed in this model. The solid blue lines represent the magnetic field, and the dashed lines correspond to the potential magnetic field. The footpoints of the two ensembles of magnetic field lines are exactly the same, and a direct comparison is meaningful.} \label{figbeta004CH}
\end{figure}

\begin{figure}[!h]
\center
{\includegraphics[width=7cm]{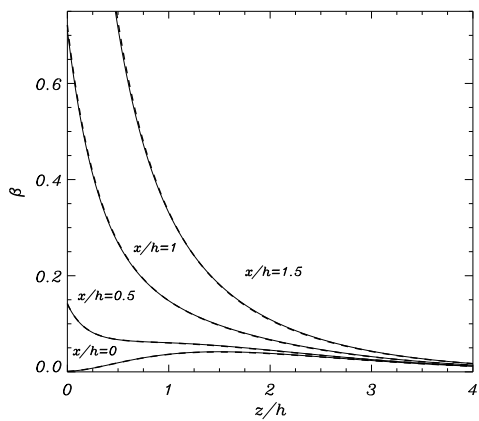}
}
\caption{\small Plasma-$\beta$ as a function of height for the CH model with $\beta_0=0.004$ at different values of $x$ (see the labels in the plot). The same parameters as in 
Fig.~\ref{figbeta004CH} are used. The thin continuous lines correspond to the results based on the semi-analytical approach, and the thick dashed lines represent the purely numerical results calculated using the code PDE2D described in Sect.~\ref{sec:numres}. The agreement between the semi-analytical and the numerical curves is quite significant.} \label{figbeta004CHbeta}
\end{figure}

To summarise, the procedure of incorporating the effect of gas pressure and gravity in the model is (a) to calculate the potential solution $A_0$ (see Sect.~\ref{sect:green}), (b) to build the function $f(A_0)$, and (c) to calculate $A_1$ using Eq.~(\ref{eq:A1double}), and (d) the final flux function is $A_0+\epsilon A_1$, where $\epsilon=\beta_0/2$ (when dimensionless variables are used). This method is applied in the following to assemble a variety of models. In Fig.~\ref{figbeta004CH} we find a constructed model for an isothermal  CH  (no temperature variation along the magnetic field lines, but temperature changes from line to line) in a low- $\beta$ situation. For the gas and temperature, we use the dependence proposed in  Eq.~(\ref{eqp0t}). We find that the plasma density is low inside the CH and grows gradually towards the coronal environment. The effect of a finite pressure and gravity changes the expansion of the magnetic field (compare the solid and dashed blue lines). The magnetic field lines approach each other in comparison with the potential magnetic field. The reason is that magnetic pressure increases to compensate for the decrease of gas pressure inside the CH and to keep the balance in the total pressure. Temperature shows a depletion inside the CH and connects smoothly with the coronal environment. Because the tempreature in this case only depends on the flux function, the temperature isocontours perfectly match the magnetic field lines, but this is not true for the density distribution. The density contrast at the centre of the CH and at $z=0$ is  $\rho_{\rm CH}/\rho_{\rm C}=(p_{\rm CH}/p_{\rm C}) (T_{\rm C}/T_{\rm CH})=0.3125$ according to the pressure and temperature ratios of this example (given in the caption of Fig.~\ref{figbeta004CH}). Because gas pressure and magnetic field change with position in the CH model, the plasma-$\beta$ is a spatially dependent function. Different cuts of this dimensionless parameter at several positions are plotted in Fig.~\ref{figbeta004CHbeta}. Now we prefer to use 1D plots to facilitate the comparison with the fully numerical results described in the following section. From Fig.~\ref{figbeta004CHbeta} we find that the plasma$-\beta$ attains its minimum values at the centre of the CH ($x=0$), where it is always below 0.05. As we move sideways from the centre, this parameter rises; the increment is especially significant at low heights. Near the position at $x/h=2,$ its value is about 6 (not shown in the figure). However, the behaviour with height of $\beta$ approaches that found at $x/h=0,$ and therefore it has low values.

\begin{figure}[h]
\center
{\includegraphics[width=7cm]{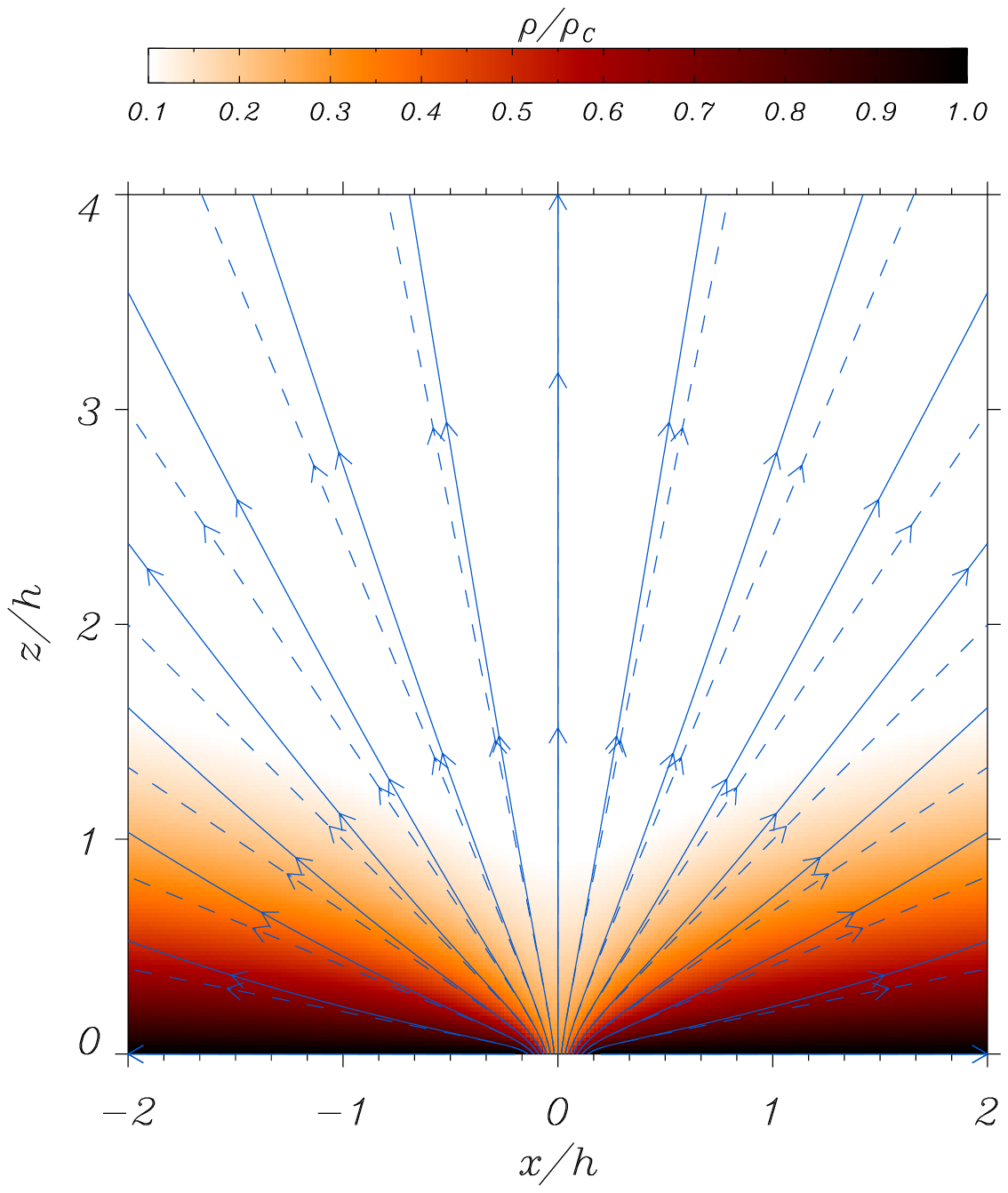}
\includegraphics[width=7cm]{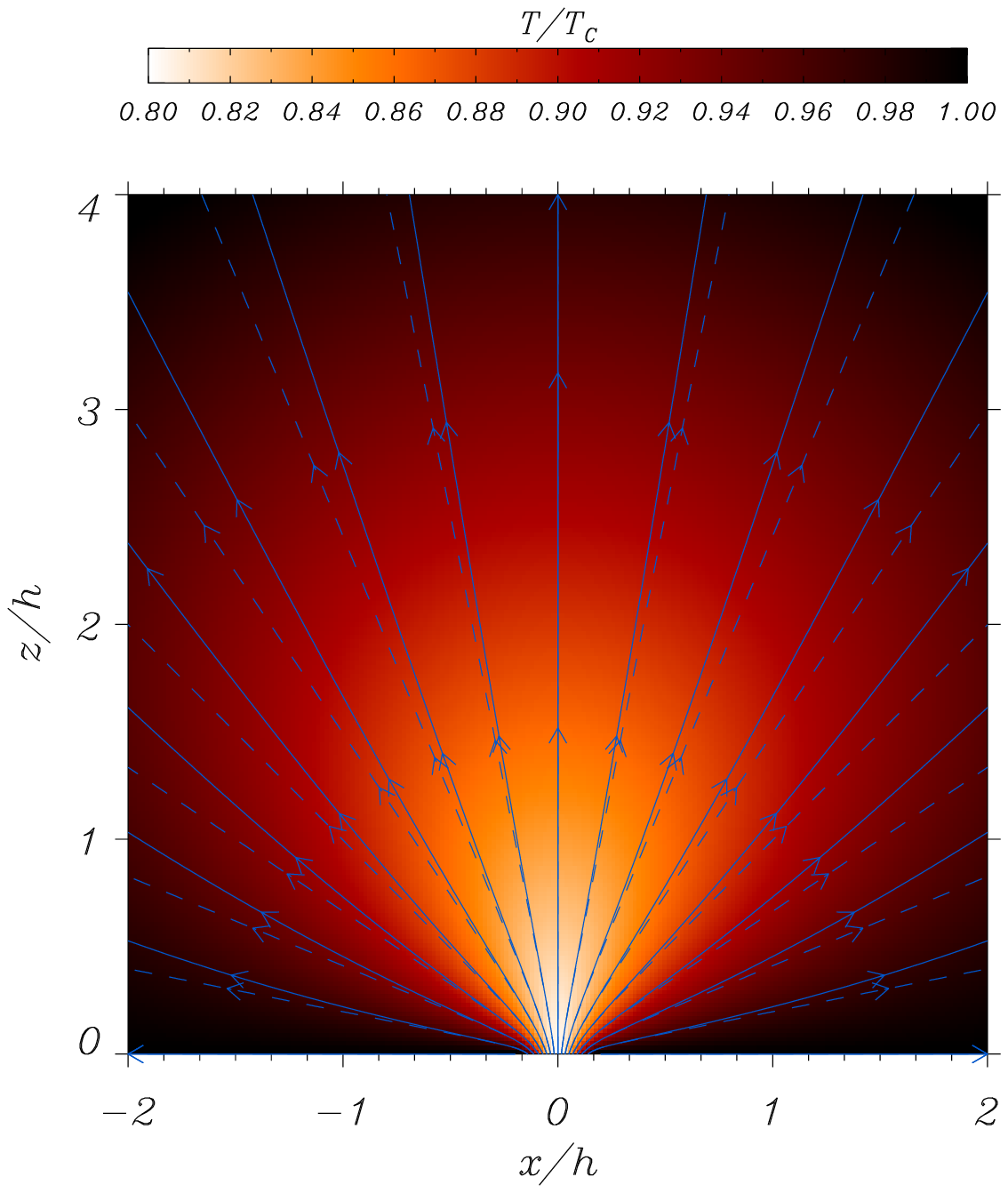}
}
\caption{\small Density and temperature distribution for the CH model with $\beta_0=0.004$. In this solution, $p_{\rm CH}/p_{\rm C}=1/4$, $T_{\rm CH}/T_{\rm C}=0.8$, and $x_0/h=0.2$. The temperature is forced to increase with height with a moderate scale height variation, $\Lambda=20 h$. The solid blue lines represent the magnetic field, and the dashed lines correspond to the potential magnetic field.} \label{figbeta004CH1}
\end{figure}

The situation with a temperature profile imposed to change with height according to Eq.~(\ref{eq:h}) is displayed in Fig.~\ref{figbeta004CH1}. Under these conditions, we find that density and the magnetic field are very similar to the previous case. The main difference is the 2D temperature distribution (and gas pressure, not shown here) in the structure. The coronal plasma surrounding the central part of the CH reaches coronal temperatures, while the lowest temperatures are found at low heights and inside the hole. This shows that in these two examples, although the density distribution is very similar, we obtain that the thermal structure of the CH model can be rather different. In any case, the specific choice given by Eq.~(\ref{eqp0t}) provides a fairly realistic representation of  typical CH conditions. It is worth mentioning that the density contrast at the centre of the CH and at $z=0$ is  the same as in Fig.~\ref{figbeta004CH} because precisely at $z=0,$ the two models have the same temperature and pressure values.

At this point and according to the results shown in Figs.~\ref{figbeta004CH} and \ref{figbeta004CH1}, it is necessary to remark that the magnetic field topology and the density structure of the CH do not coincide. This has important implications regarding the setup and interpretation of MHD simulations of CHs, but also for the interpretation of the observations of CHs and the possible location of the CHB. Nevertheless, our model does not provide information about plumes and rays in CHs, which are nearly radially aligned density striations that are thought to follow the ambient magnetic field. 

\begin{figure}[h]
\center
{\includegraphics[width=7cm]{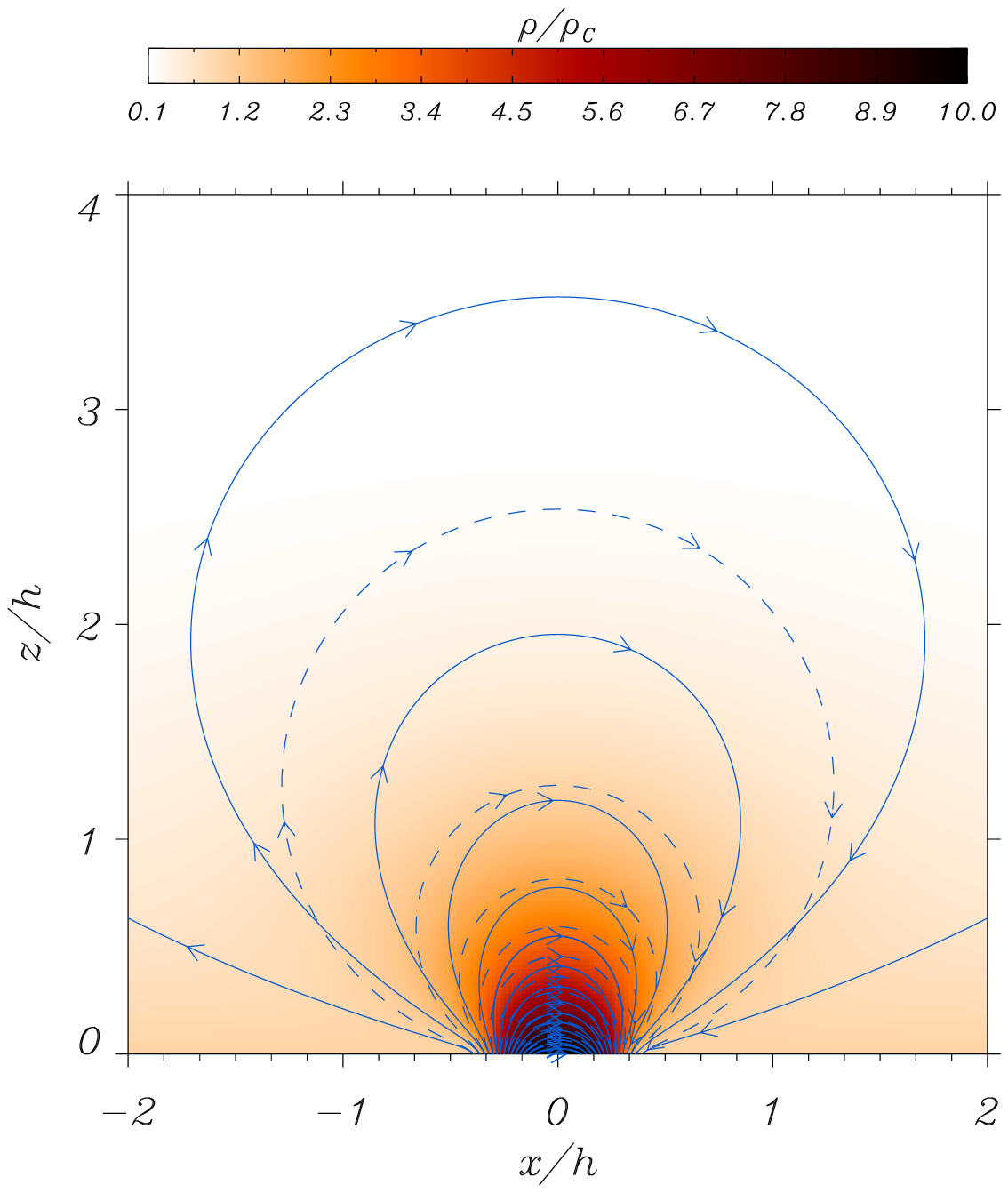}
\includegraphics[width=7cm]{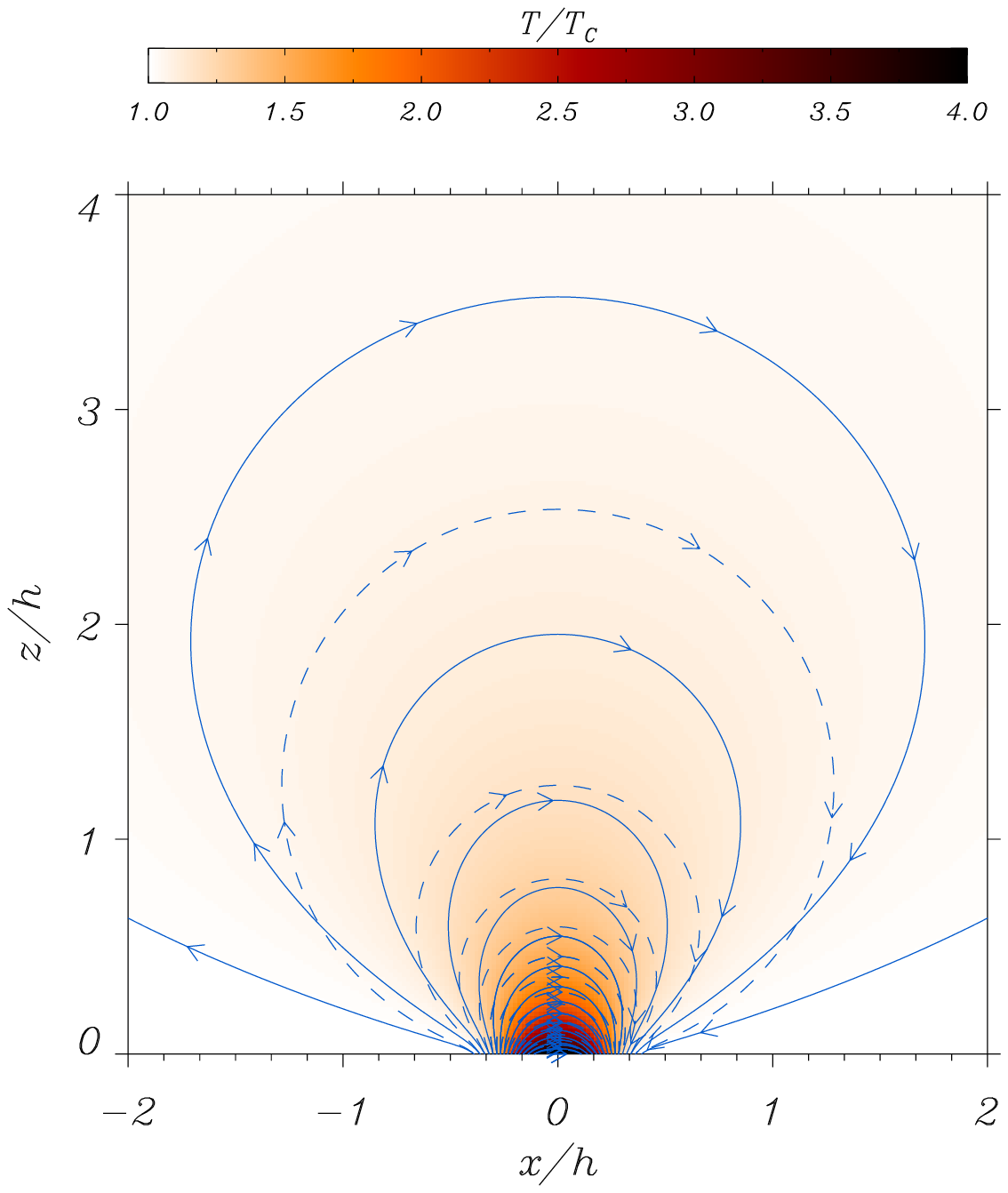}
}\caption{\small Density and temperature distribution for the AR model with $\beta_0=0.004$. In this solution, $p_{\rm AR}/p_{\rm C}=40$, $T_{\rm AR}/T_{\rm C}=4$, $x_0/h=0.2$, $x_1/h=-1/4$, and $x_2/h=1/4$. The isothermal condition (along the field lines) is imposed in this model. The solid blue lines represent the magnetic field, and the dashed lines correspond to the potential magnetic field.}
 \label{figbeta004AR0}
\end{figure}

We now concentrate on ARs and compose several equilibrium models using the same semi-analytical procedure and the functional dependences given by Eq.~(\ref{eq:arprof}).  Figure~\ref{figbeta004AR0} shows the configuration that we obtain when $\mathcal{H}(z)=1$. Density is strongly localised at the core of the AR, and gravity produces the stratification effect that is visible outside the core. The magnetic field lines expand much more than in the potential case. This is essentially produced by the high-density and high-pressure core. In this case, magnetic pressure has to decrease to balance the excess of gas pressure at the core, and this produces a strong separation of the field lines compared to the potential case. Temperature is also higher at the core, reaching 4 MK, than for the coronal temperature at 1 MK. These two temperatures are imposed in the model through the parameters $T_{\rm AR}$ and $T_{\rm C}$. The density contrast at the centre of the AR and at $z=0$ is  $\rho_{\rm AR}/\rho_{\rm C}=(p_{\rm AR}/p_{\rm C}) (T_{\rm C}/T_{\rm AR})=10$ using the pressure and temperature ratios of the model (given in the caption of Fig.~\ref{figbeta004AR0}).

\begin{figure}[h]
\center
{\includegraphics[width=7cm]{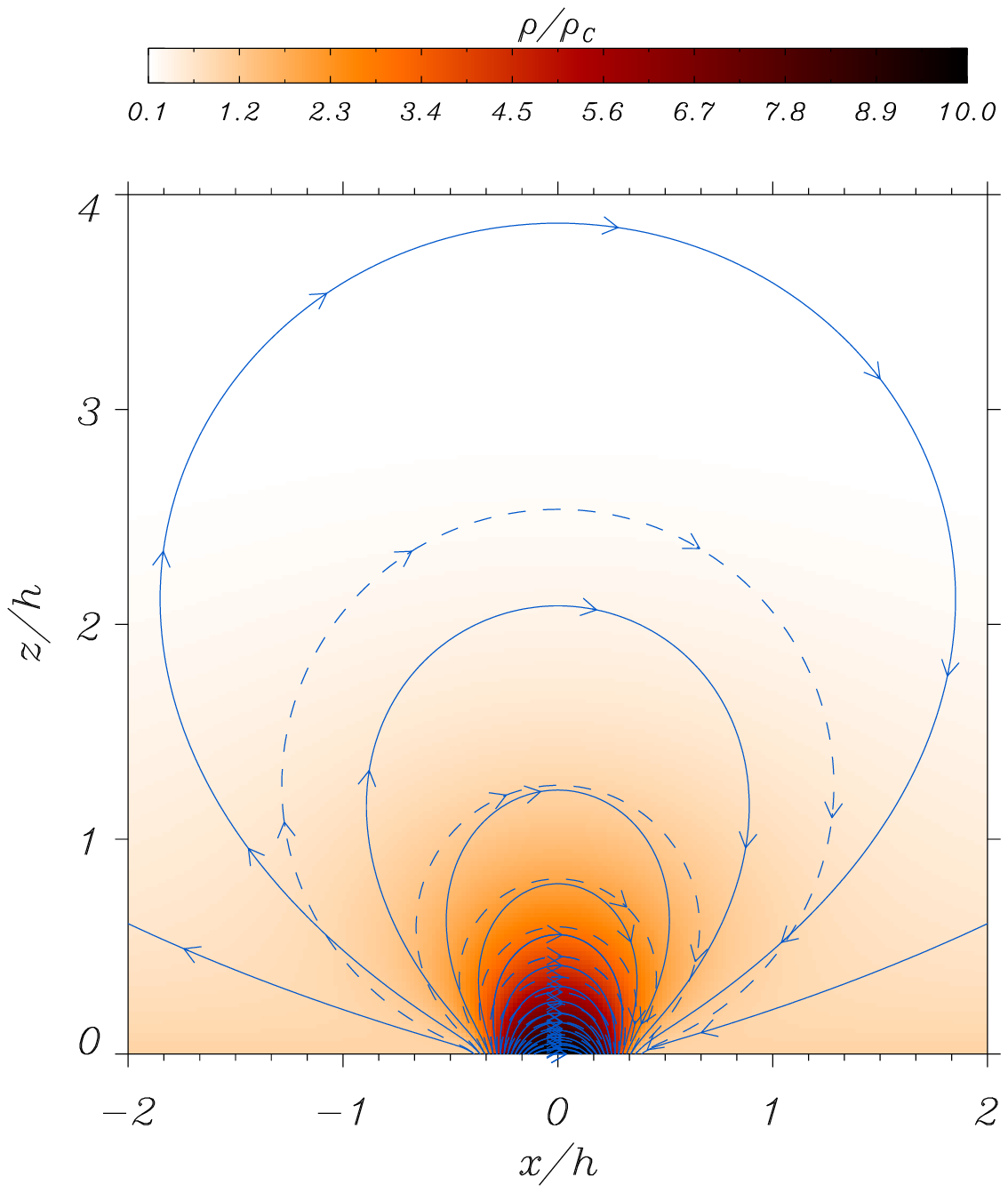}
\includegraphics[width=7cm]{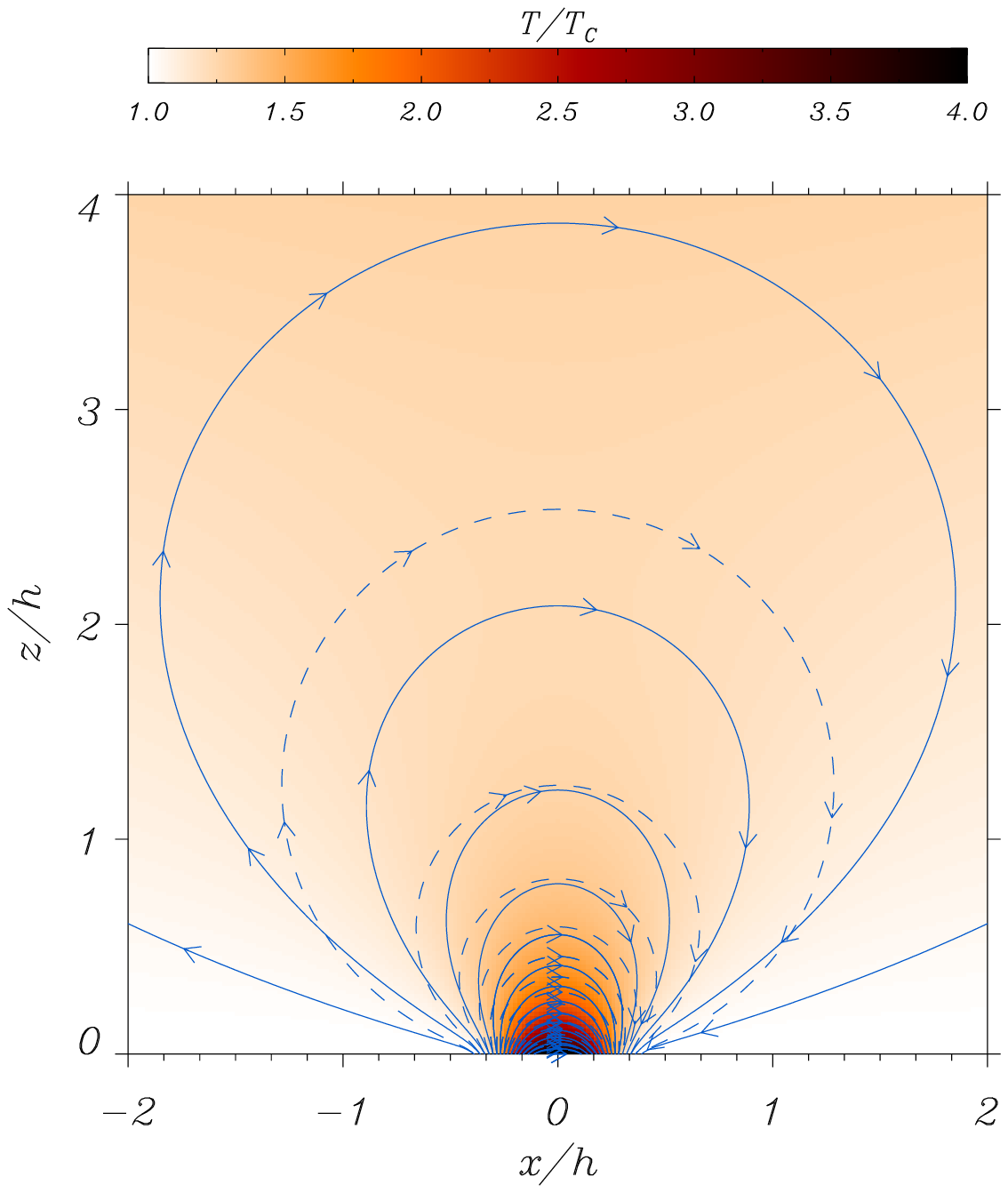}
}
\caption{\small Density and temperature distribution for the AR model with $\beta_0=0.004$. In this solution, $p_{\rm AR}/p_{\rm C}=40$, $T_{\rm AR}/T_{\rm C}=4$, $x_0/h=0.2$, $x_1/h=-1/4$, and $x_2/h=1/4$. The temperature is forced to increase with height with a moderate scale height variation, $\Lambda=20 h$.}
 \label{figbeta004AR1}
\end{figure}

The situation with a temperature increasing with height, $\Lambda=20 h$, is shown in Fig.~\ref{figbeta004AR1}. The density distribution is similar to the previous case, and the main differences are in the temperature distribution, which is higher in the plasma surrounding the core. Because in this configuration the temperature changes along the field lines, the thermal conduction has an effect; this is discussed in Sect.~\ref{sect:energy}. The temperature dependence on height has some influence on the structure of the magnetic field as (cf. Fig.~\ref{figbeta004AR0}). Equation~(\ref{eq:arprof}) provides a route for building AR models that are similar to the diffuse background of bipolar regions reported in the observations. The structure of individual loops commonly found in real ARs is missing in our model and is considered as secondary for our purposes.

\subsection{Numerical calculation of the equilibria in the low-$\beta$ regime}\label{sec:numres}

When the plasma-$\beta$ is not small, the previous semi-analytical method we used to find a solution is not fully justified. In this case, it is convenient to solve Eq.~(\ref{eqgradshastatic}) by numerical means. Even in the low-$\beta$ regime, it is useful to have an alternative method for a comparison with the results of Sect.~\ref{sec:semi-an}.  We used the numerical code PDE2D \citep{Sewell2018}, which uses finite elements to solve partial differential equations of the type found in our problem. We chose a collocation method with a bicubic Hermite basis functions. This choice of basis functions ensures that the first derivatives of the approximate solution are all continuous. Newton's method was used to iteratively solve the nonlinear algebraic equations resulting from the collocation method. Convergence is in general achieved in just four or five iterations when a given constant value as the initial guess of the solution in the whole numerical spatial domain is chosen \citep[see][for another alternative numerical technique]{Pizzo1986}.  

%We find a considerably good  performance of the selected %numerical method. 

The particular choice of boundary conditions discussed earlier was also introduced into the numerical scheme and constitute an essential part of the problem. We wished to minimise the effect of the lateral and upper boundaries on the solution, and as we have shown, a convenient approach is to impose the BC at infinity, except at $z=0,$ where we selected the particular profile for the vertical component of the magnetic field. This was achieved by choosing a suitable coordinate system. In  the $z-$direction, we used the following transformation, sometimes referred to as a M\"obius transformation:\begin{align}
Z=\frac{1}{z/L_z+1},
\end{align}
and the initial domain $0< z < \infty$ changes to $0< Z < 1,$ which is bounded. There is some freedom in choosing the factor $L_z$ in the transformation. In the $x-$direction, we can choose an equivalent transformation if the system is symmetric with respect to $x=0$. Nevertheless, we preferred not to be restricted to this situation. For this reason, our choice was
\begin{align}
X=\tanh\left(x/L_x\right),
\end{align}
and therefore, instead of solving the problem in the range $-\infty< x < \infty$ with the new coordinate, we just need to consider the interval $-1< X < 1$. Again, $L_x$ is a factor that we have to select.

With the previous coordinate transformations, we rewrite the 2D Laplacian in terms of the new coordinates by using the corresponding scale factors. We obtain
\begin{align}
 \frac{\partial^2 A}{{\partial x^2}}+\frac{\partial^2 A}{{\partial z^2}}=&\frac{1-X^2}{L_x^2}\left(-2 X \frac{\partial A}{\partial X}+(1-X^2)\frac{\partial^2 A}{\partial X^2}\right)\nonumber \\&+\frac{Z^2}{L_z^2}\left(2 Z \frac{\partial A}{\partial Z}+Z^2\frac{\partial^2 A}{\partial Z^2}\right). \label{eq:gradshacoord}
\end{align}
Due to the coordinate transformation, first-order derivatives of $A$ are present now. This equation plus the pressure term given by Eqs.~(\ref{eqderivP}) and (\ref{eqderivPnoiso}) was implemented in the numerical code PDE2D for the specific functions defined in Eqs.~(\ref{eqp0t}) or (\ref{eq:arprof}). After we obtained a solution in the coordinates $X$ and $Z,$ we mapped to Cartesian coordinates to better visualise the results. The obtained solution in Cartesian coordinates must be independent of the factors $L_x$ and $L_z$ used in the transformation. This provides clues about the optimal choice of these parameters because in essence, they control the spacing of the non-uniform grid viewed in Cartesian coordinates. For example, for the AR model, the parameter $L_x$ must be of the order of the separation of the two unipolar regions, $2 x_1$. 

The numerical results for exactly the same situation as in Fig.~\ref{figbeta004CH} (calculated using the semi-analytical approach) were computed numerically using a grid of $200\times200$ points. A comparison of the plasma-$\beta$ using the two methods is available in Fig.~\ref{figbeta004CHbeta}. This figure demonstrates that the two results are almost identical and that the differences appear when the plasma$-\beta$ rises. This corroborates that the semi-analytical method is accurate enough to be used as a tool to formulate equilibria in the low-$\beta$ regime. Adding second-order terms to the semi-analytical approach would increase the accuracy of the method further.

Another procedure for verifying that the semi-analytical and/or the numerical solutions are correct is to introduce the obtained profile for $A(x,z)$ and the corresponding pressure and density distributions in Eq.~(\ref{eqgradshastatic}) and to compute the force balance. For the purely numerical results, we find that as the resolution is increased, the force balance converges towards zero, indicating that our numerical solutions are truthful representations of the real solution to Eq.~(\ref{eqgradshastatic}). Nevertheless, we have found that as the plasma-$\beta$ rises (typically for $\beta_0=0.1$), the numerical solution is not as accurate, and convergence issues appear. These issues are not mitigated when the spatial resolution is increased, and they are more likely related to the intrinsic nature of the nonlinear terms.

\section{Energy balance}\label{sect:energy}

In Sect.~\ref{sec:semi-an} we have obtained magnetohysdrostatic solutions for a balance between the Lorenz force, the pressure gradient, and the gravity force. In addition to the force equilibrium, the system should also satisfy energy balance. When we consider that we are in ideal MHD, the response is trivial because the effects of thermal conduction, radiative losses, and heating are ignored. In this case, the equilibrium solutions derived in the previous sections under force balance can be used together with an adiabatic energy equation to study different processes, for example,  the temporal evolution of MHD waves in the system. A possible application of the models developed in the present work is the analysis of the evolution of global MHD waves interacting with CH or AR because in this problem, the assumption of ideal MHD is justified as a first step if the focus is on the properties of the waves.

In reality, the assumption of ideal MHD in the solar corona is a poor approximation because thermal conduction, optically thin radiation, and heating are processes that cannot be ignored in general (it depends on the temporal scales we are interested in). We need to address how these effects are included in our models. The question is in fact simple and clearly explained in \citet{low1975,low1980}, the energy equilibrium (or thermal balance) can be calculated only a posteriori in the scheme used in this paper. When we have the force balanced solution, that is, the temperature and density distribution in 2D, we can calculate the conduction term, $E_{\rm C}$, and the radiation term, $E_{\rm R}$, according to their expressions and then derive the distribution of the heating term, $H$, to have a perfect energy balance. In this situation, we derive the heating distribution from the model using the following equation:
\begin{align}
    H(x,z)=E_{\rm C}(x,z)+E_{\rm R}(x,z).\label{eq:heat}
\end{align}
With this procedure, we do not calculate the temperature distribution self-consistently according to a known energy equation. This can only be achieved, and this point is relevant, assuming an ad hoc explicit dependence of the heating function, for example, with the density, temperature, or magnetic field. The exact form of the heating function and its source is precisely one of the main unknowns in relation to the coronal heating problem, however. Nevertheless, deriving the heating distribution after we obtain a force balance for a prescribed temperature and pressure profiles is an approach that is worthwhile investigating. The available energy in the system is in some way constrained by the requirement of force balance. This does not necessarily mean that the heating function obtained using this method is closely related to the real heating source in the solar corona, but it can still provide useful information about the energetics of system. In the following, we try to make this point clearer. As far as we know, this distinctive approach has not been explored in the literature, at least in the form proposed here in relation to the 2D problem.

%This was achieved by \cite{lowetal2012} under solar prominence %conditions.

We start with the most elementary situation, the model of constant temperature along the magnetic field lines. Temperature can change from line to line, however, due to the dependence $\mathcal T(A)$. Thermal conduction is zero under this assumption (we did not consider conduction perpendicular to the magnetic field). This means that the only possibility for thermal equilibrium is that the heating has to balance the radiative losses exactly. This is a rather unlikely situation from the physical point of view, but it has been suggested by \citet{aschwandenetal1999a,aschwandenetal2000a} because their observations indicate that coronal loops of different ARs are essentially in an isothermal state, and therefore thermal conduction is essentially zero. Nevertheless, the focus of our work is not coronal loops, but the diffuse background, which does not necessarily need to have the same isothermal character.

%These authors concluded that the observed loops are most likely not %in hydrostatic equilibrium, i.e., the static assumptions of the model %are too far from a real situation in which loops are in a sort of %dynamic state. Nevertheless, the observations of %\citet{warrenetal2012, warrenetal2020} indicate that the hot %components of AR are essentially in a quasi steady state. 

Thermal conduction has an effect in our model if we consider that temperature also depends on height, that is, $T(A,z)$.  We know that the thermal conduction term in the energy equation \citep[see e.g. ][]{priest1982} can be expressed as 
\begin{align}\label{eq:Ec}
E_{\rm C}=\nabla \cdot {\bf q}=-\nabla \cdot \bigg(\kappa_\parallel \left(\nabla T \cdot {\bf \hat{B}}\right){\bf \hat{B}}\bigg),
\end{align}
which is written here in terms of the unitary magnetic field vector ${\bf \hat{B}}={\bf B}/B$. The heat flux is represented by $\bf q,$ and it is parallel to the magnetic field (assuming that $\kappa_{\perp}=0$, as mentioned earlier). Using vector identities and the fact that $\nabla \cdot {\bf B}=0,$ the previous expression reduces to 
\begin{align}
E_{\rm C}=-{\bf B}\cdot \nabla\left(\frac{\kappa_\parallel}{B^2} \left(\nabla T \cdot {\bf B}\right)\right)=-\frac{d}{dl}\left(\kappa_\parallel \frac{d T}{dl}  \right)+\frac{\kappa_\parallel}{B}\frac{dB}{dl} \, \frac{dT}{dl},\label{eq:condsimp}
\end{align}
where on the right-hand side we used the coordinate $l$ along the magnetic field {\bf B}. Thermal conduction has two contributions, related to the variation of temperature along the field lines and also to changes in the modulus of the magnetic field along $l,$ or equivalently, to the change in area of the corresponding flux tube. The conduction term, $E_{\rm C}$, can be either positive or negative and depends on the location in the domain, hence it is a space-dependent function.

Using the separable form for the assumed temperature dependence in this work,  Eq.~(\ref{eq:Tsep}), we evaluated the factors in the heat flux definition, finding that 
\begin{align}
\nabla T \cdot {\bf B}&=\frac{\partial T}{\partial x} B_x+ \frac{\partial T}{\partial z} B_z\nonumber \\
&= \frac{\partial \mathcal {T}}{\partial A} B_z B_x \mathcal{H}-\frac{\partial \mathcal{T}}{\partial A} B_x B_z \mathcal{H}+\mathcal{T} B_z \frac{d \mathcal{H}}{dz}
=\mathcal{T} B_z \frac{d \mathcal{H}}{dz},
\end{align}
where we used the chain rule and the definition of the magnetic field components in terms of the flux function $A$. Using the previous expression and the fact that $\kappa_\parallel=\kappa_0\, T^{5/2}$ ($\kappa_0=1.1\times 10^{-11}$ W m$^{-1}$ K$^{-7/2}$), the heat flux vector reads
\begin{align}
{\bf q}=-\kappa_0 \mathcal{T}^{7/2}\frac{B_z}{B^2} \frac{d \mathcal{H}}{dz} \mathcal{H}^{5/2}\, {\bf B}.
\end{align}

\noindent For the explicit exponential dependence of the function $\mathcal{H}(z)$ given by Eq.~(\ref{eq:h}), we obtain that the vertical component of the heat flux vector is
\begin{align}
q_z={\bf q}\cdot{\bf \hat{e}_z}=-\kappa_0 \mathcal{T}^{7/2}\frac{B^2_z}{B^2} e^{(7/2)\, z/ \Lambda}\frac{1}{\Lambda}.\label{eq:heatB}
\end{align}
When the temperature is independent of $z,$ the parameter $\Lambda$ tends to infinity and the heat flux is zero because we are in the isothermal situation along each field line. Equation~(\ref{eq:heatB}) indicates that a finite $\Lambda$ introduces a net heat flux in the vertical direction. At the bottom layer of the domain ($z=0$), we have an incoming vertical heat flux for negative $\Lambda,$ while it is outgoing for positive $\Lambda$. This has a relevant effect  because heat, coming from below our reference level for positive $\Lambda$, modifies the thermal balance of the system. The opposite situation, that is, heat leaving the system through the bottom boundary, is only possible if the temperature increases with height.  The vertical  heat flux is independent of the sign of $B_z$ according to Eq.~(\ref{eq:heatB}). 

The horizontal heat flux is
\begin{align}
q_x={\bf q}\cdot{\bf \hat{e}_x}=-\kappa_0 \mathcal{T}^{7/2}\frac{B_z B_x}{B^2} e^{(7/2)\, z/ \Lambda}\frac{1}{\Lambda}.\label{eq:heatBx}
\end{align}
In contrast to the vertical component, the horizontal component of the heat flux depends on the signs of $B_z$ and $B_x$ , and this is a consequence of the fact that the heat flux vector points along the magnetic field. Only in the situation of a purely vertical magnetic field (the centre of a symmetric CH) or a purely horizontal magnetic field (the centre of a symmetric AR) is the horizontal heat flux zero. It is worth mentioning that even in the situation of zero heat flux, thermal conduction can be different from zero (because this magnitude is the divergence of the heat flux, Eq.~(\ref{eq:Ec})).

From the previous expressions, it is easy to obtain the explicit form of the energy conduction when the magnetic field is purely vertical or purely  horizontal. For the purely  vertical magnetic field at the centre of the CH, we find (see Eq.~(\ref{eq:condsimp})) that 
\begin{align}
    E_{\rm C}(x=0,z)=&-\frac{\partial}{\partial z}\left(\kappa_0 \mathcal{T}^{7/2}\frac{d \mathcal{H}}{dz} \mathcal{H}^{5/2}\right)\nonumber \\ &+\kappa_0 \mathcal{T}^{7/2}\frac{d \mathcal{H}}{dz} \mathcal{H}^{5/2} \frac{\partial |B_z|}{\partial z}\frac{1}{|B_z|}.\label{eq:EcCH}
\end{align}
This expression applied to the exponential dependence on height of the temperature reduces to
\begin{align}
    E_{\rm C}(x=0,z)&=\kappa_0 \mathcal{T}^{7/2} e^{(7/2)\, z/ \Lambda}\frac{1}{\Lambda}
    \left(-\frac{7}{2}\frac{1}{\Lambda}+\frac{\partial |B_z|}{\partial z}\frac{1}{|B_z|}\right).\label{eq:EcCHlam}
\end{align}
For the CH model we used, we have that $\partial |B_z|/\partial z<0$ at $x=0$, regardless of $z$. This means that for $\Lambda>0,$ we always obtain that $E_{\rm C}(x=0,z)<0$ according to Eq.~(\ref{eq:EcCHlam}). For $\Lambda<0, $ however, the sign of the energy conduction can change depending on the terms inside the parentheses. In particular when 
\begin{align}
\frac{7}{2}\frac{1}{|\Lambda|}<\left|\frac{\partial |B_z|}{\partial z}\right|\frac{1}{|B_z|},\label{eq:condheat}
\end{align}
then $E_{\rm C}(x=0,z)>0$. Nevertheless, this condition might be only satisfied up to a certain height, as we show below.

For a purely horizontal magnetic field at the centre of the symmetric AR model, it is possible to perform a similar analysis as for the CH at $x=0$. In this case, it is easy to show that 
\begin{align}
    E_{\rm C}(x=0,z)&=-\kappa_0 \mathcal{T}^{7/2} e^{(7/2)\, z/ \Lambda}\frac{1}{\Lambda}
    \frac{\partial B_z}{\partial x}\frac{1}{B_x},\label{eq:EcArlam}
\end{align}
which is different from Eq.~(\ref{eq:EcCHlam}). 
Because the  magnetic configuration of the AR has a concave geometry at $x=0,$  $(\partial B_z/\partial x)/B_x<0$. Hence, for $\Lambda>0$  Eq.~(\ref{eq:EcArlam}) indicates that we always obtain that $E_{\rm C}(x=0,z)>0$, while for $\Lambda<0,$ we have the opposite situation, $E_{\rm C}(x=0,z)<0$. These characteristics of the sign of the conduction term have important consequences for the energy balance. 

\begin{figure}[!hh]
\center
{\includegraphics[width=6.5cm]{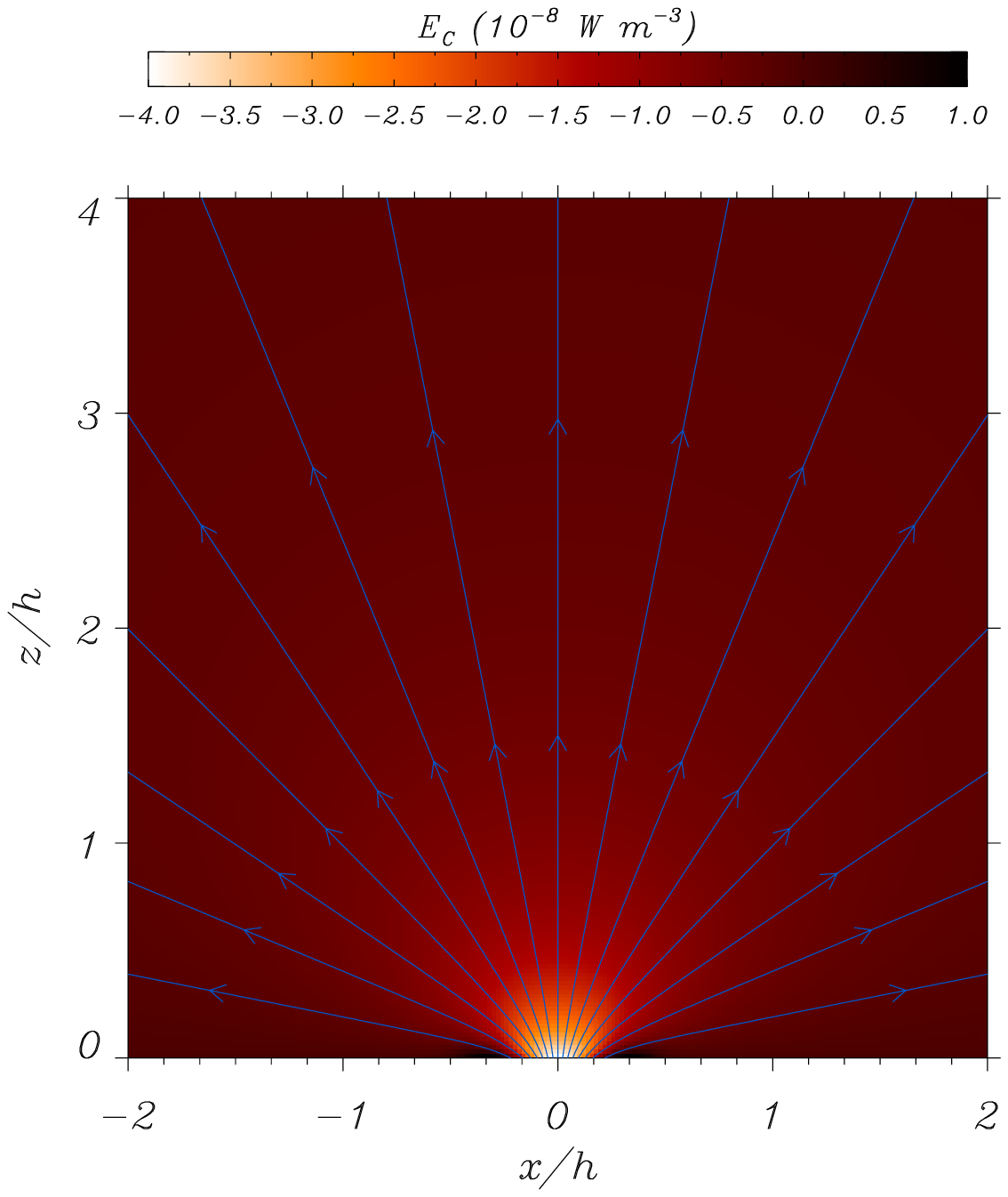}
\includegraphics[width=6.5cm]{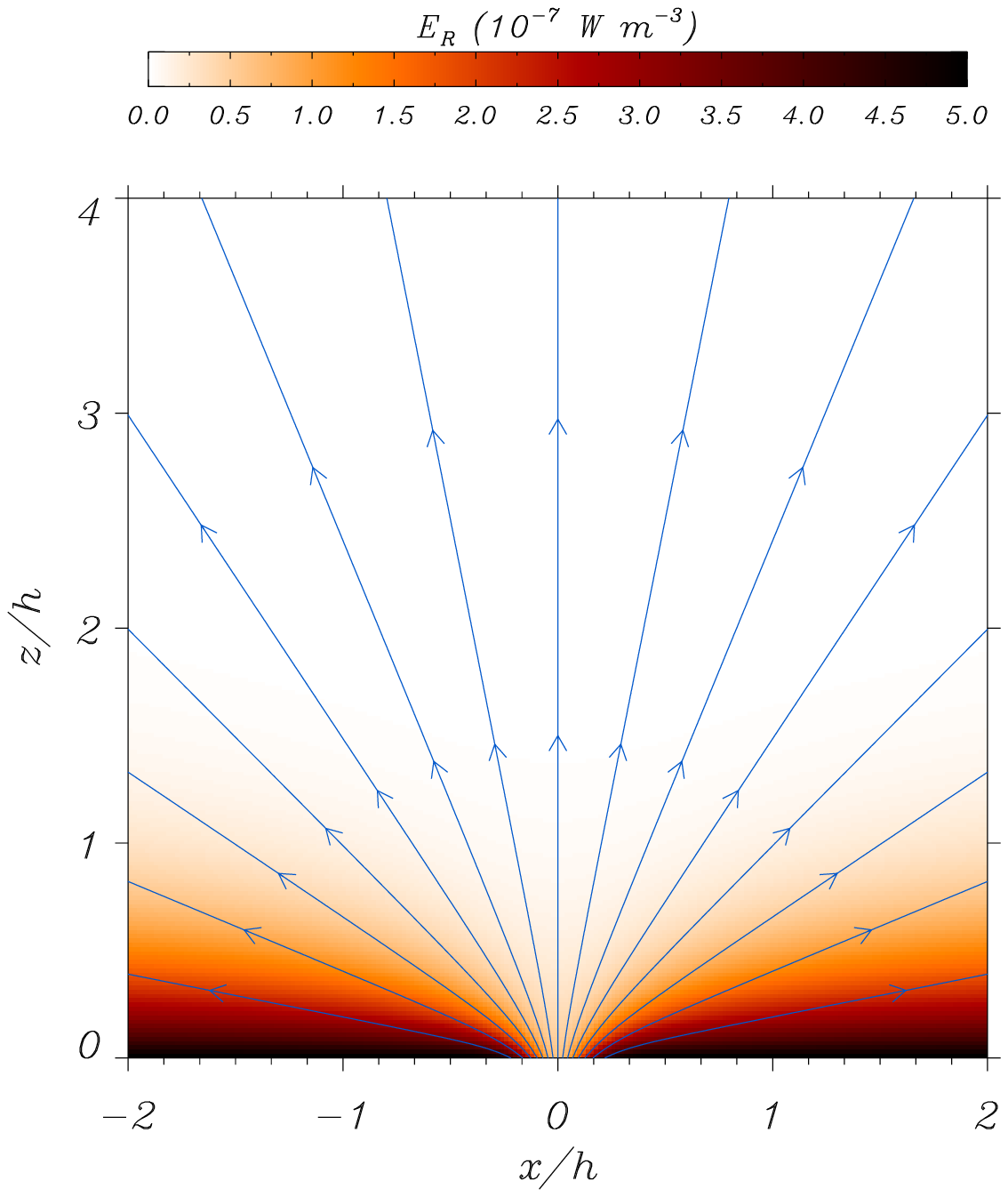}
\includegraphics[width=6.5cm]{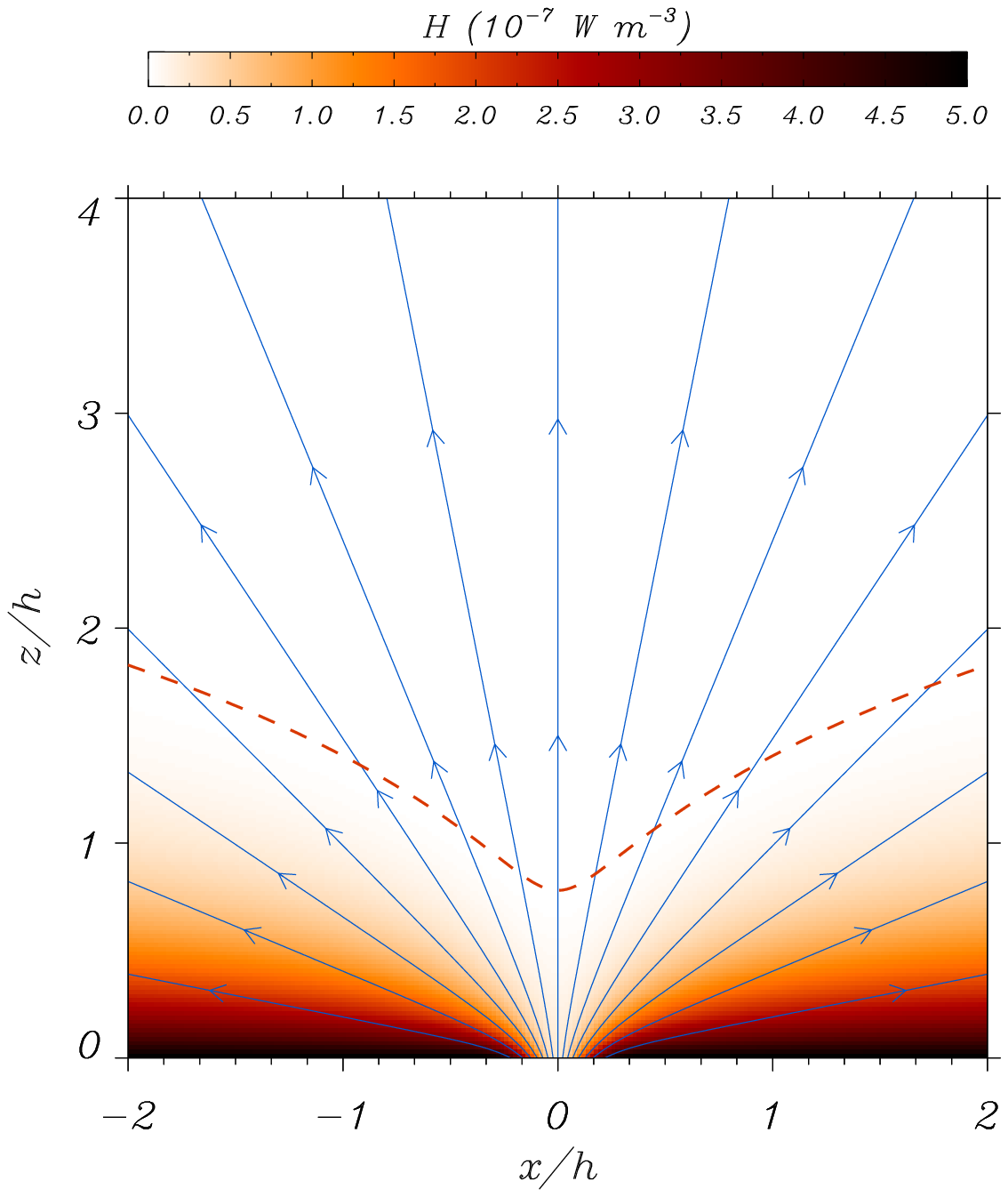}
}
\caption{\small Conduction, radiation, and heating terms for the CH model. The boundary marked with dashed lines in the heating term separates the transition from energy balance to non-energy balance (energy balance below the curve, non-energy balance above the curve). In this plot, $\Lambda=200h$. In our model, the reference values are $\rho_0=5\times 10^{-13}\rm kg\, m^{-3}$ and $T_0=1\,\rm MK$.} \label{figenergyterms}
\end{figure}

We studied the radiative losses. Because we considered coronal plasmas, we used the optically thin losses of \citet{hildner1974} for simplicity, other functions can be used, however (see \citealt{athay1986}, \citealt{dereetal1997}, \citealt{klimchukcarg2001}, and \citealt{landietal2012}). As usual, $E_{\rm R}=\rho^2 Q(T)$, where $Q(T)$ is the corresponding loss function depending only on the temperature. From the force balance equation, we obtained the density and the temperature distribution in 2D, and this information was used to compute the spatial distribution of the 2D radiative losses using the previous expression. When we know the radiative losses and the conduction term, it is straightforward to calculate the spatial distribution of the heating to have energy balance using Eq. (45). Based on physical grounds, the heating must be positive, representing a source of energy that needs to be supplied for the system to have energy balance. Nevertheless, in our approach, it may happen that in some regions of the 2D configuration, the condition of positiveness is not satisfied. The reason is that as we showed above, the conduction term is either positive or negative, while the radiation term is always positive. The sum of these two terms is not necessarily a positive number. In this case, the  obtained heating is negative and represents a nonphysical energy sink. The system cannot achieve energy balance in this situation. We demonstrate this behaviour by calculating the different terms in the energy equation for several of the models presented earlier. The results for a CH are shown in Fig.~\ref{figenergyterms}. The conduction term is negative in most of the domain and is typically one order of magnitude smaller that the radiation term. In particular, we showed that for the vertical magnetic field line at the centre of the CH, the conduction term is always negative when the temperature increases with height (the present example). In this situation, the sum of radiation and conduction is  positive up to one scale height at $x=0$. The dashed line in the plot for the heating $H$ represents the curve that separates the transition between the energy-balanced and non-energy-balanced points of the domain. At low heights, typically below $z=h$, the heating has to balance the radiation losses, and it has a rather reduced value inside the CH, where density and temperature are lower than in the environment. In this example, the gradient of the temperature with height is quite weak because $\Lambda=200h$, but this has a relevant effect due to the profile of conduction term.

\begin{figure}[!h]
\center
{\includegraphics[width=7.cm]{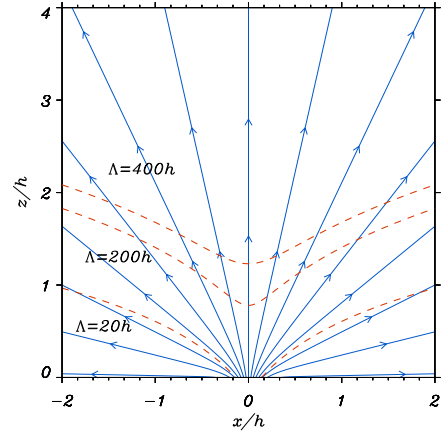}
}
\caption{\small Example of the location of the boundary, dashed red lines, that separate regions in which energy balance is allowed (below these lines) from regions in which it is not permitted (above the lines). In this case, the temperature increases with height ($\Lambda>0$). The magnetic field is also displayed with blue lines.} \label{figCHlambda}
\end{figure}

The location of the transition curve depends on the parameters, as Fig.~\ref{figCHlambda} indicates. For example, for $\Lambda=20h$, no energy balance is possible inside the CH even at $z=0$. Therefore, this situation is difficult to achieve from the physical point of view. Nevertheless, when $\Lambda$ rises, see $\Lambda=200h$ and $400h$, the boundary moves progressively upward, meaning than in the range $0<z<1.5 h$ at $x=0$, energy balance is accomplished for $\Lambda=400h$. A small vertical gradient in temperature has a significant effect on the overall energy balance in the system.

\begin{figure}[!h]
\center
{\includegraphics[width=7.cm]{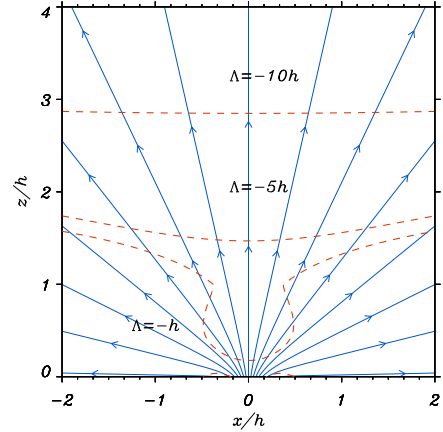}
}
\caption{\small Same as in Fig.~\ref{figCHlambda}, but the temperature decreases with height ($\Lambda<0$).} \label{figCHlambda1}
\end{figure}

\begin{figure}[!h]
\center
{\includegraphics[width=6.5cm]{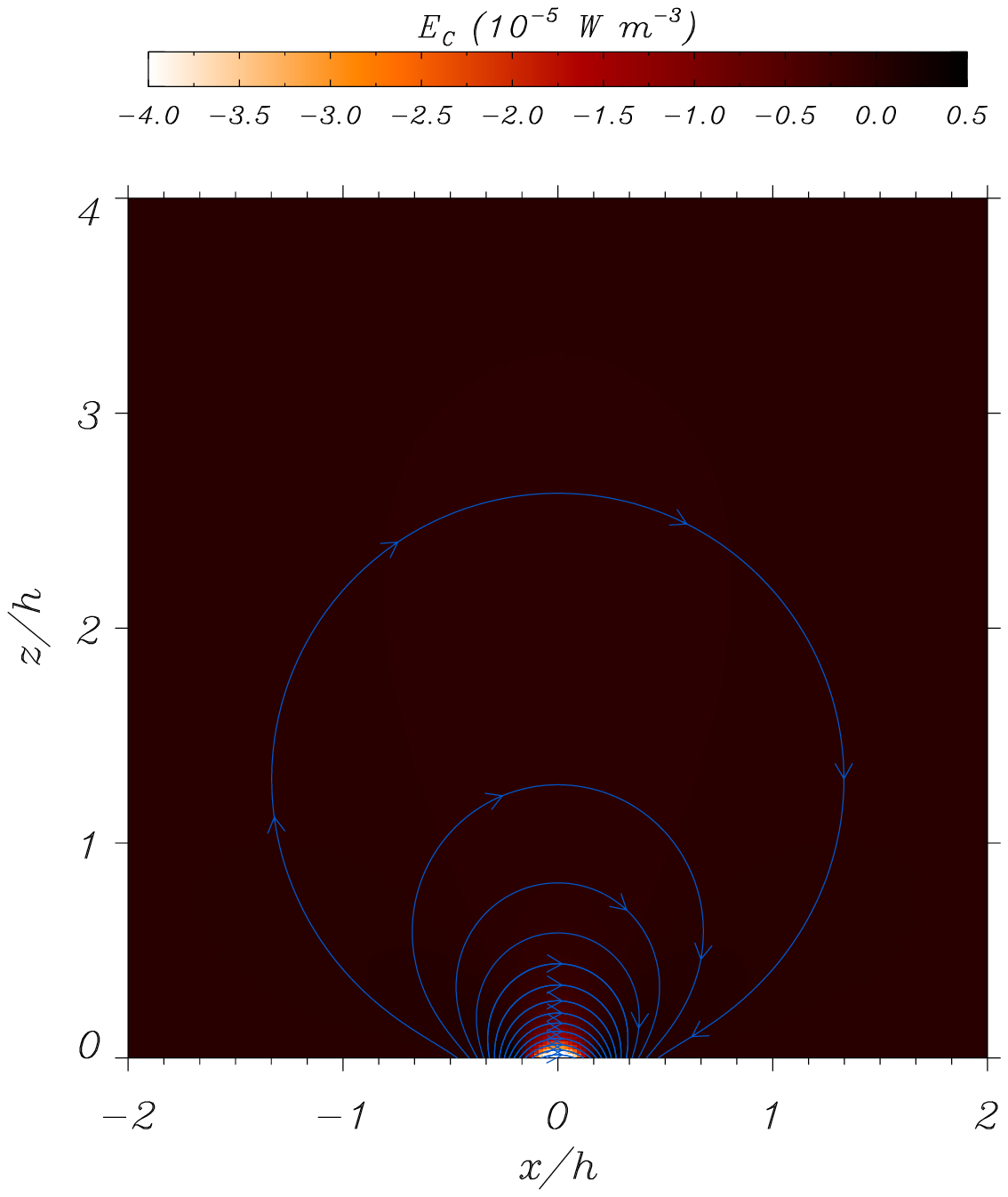}
\includegraphics[width=6.5cm]{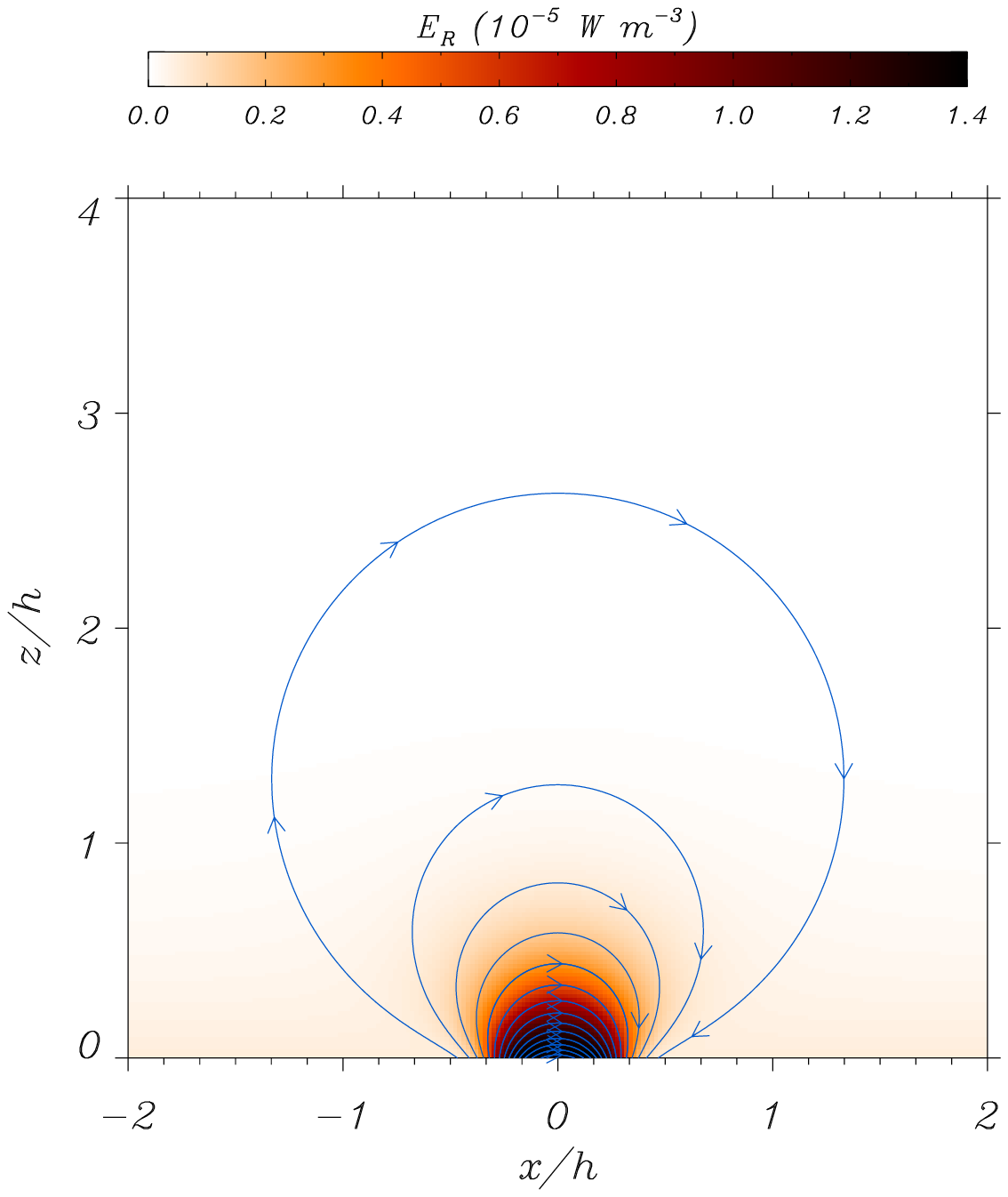}
\includegraphics[width=6.5cm]{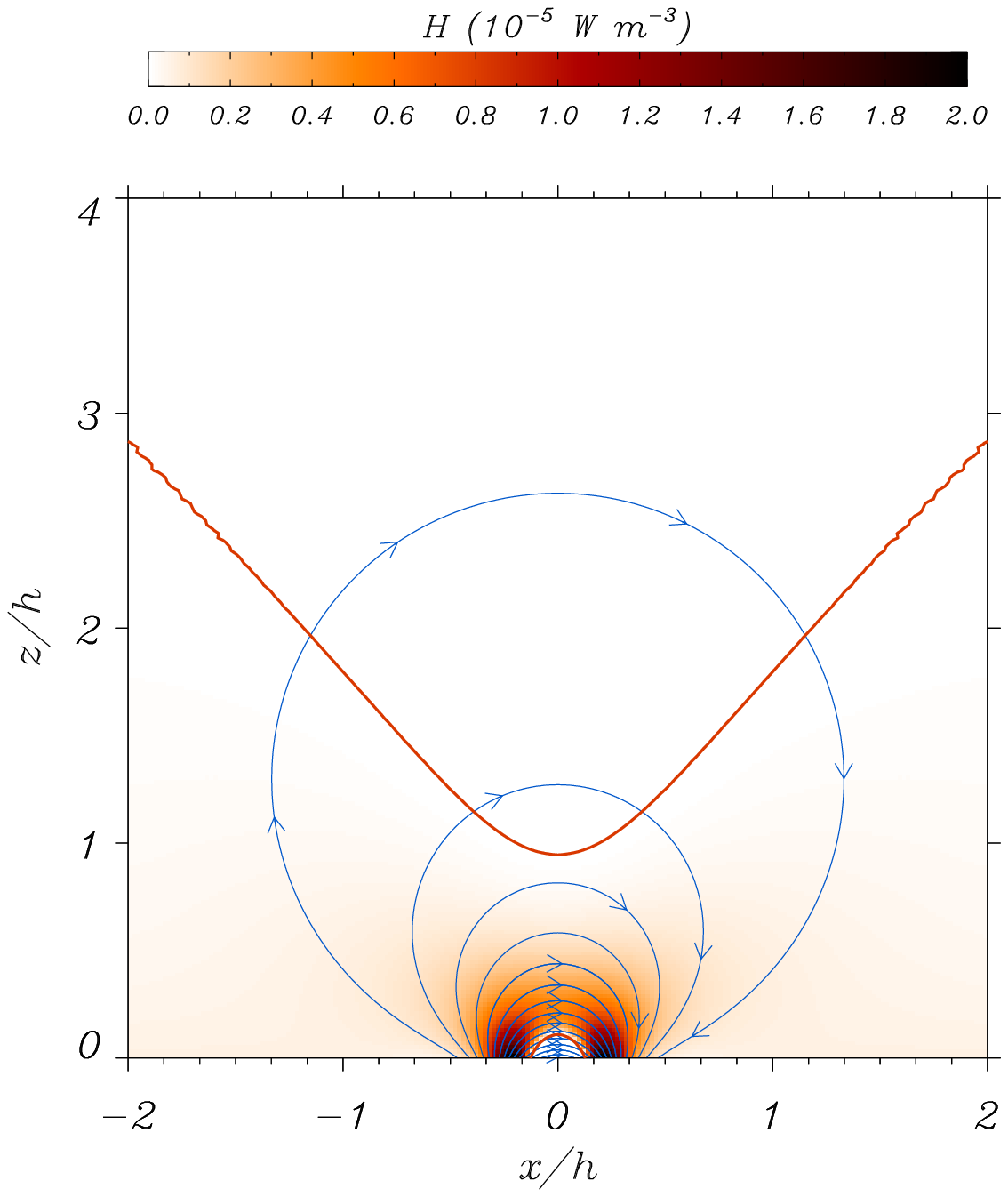}
}
\caption{\small Conduction, radiation, and heating terms for the AR model. The boundary marked with a continuous line in the heating term separates the transition from energy balance to non-energy balance (energy balance below the curve, non-energy balance above the curve). In this plot, $\Lambda=-20h$.} \label{figenergytermsAR}
\end{figure}

Figure~\ref{figCHlambda1} shows the same type of plot as in Fig.~\ref{figCHlambda},  but for a  decreasing temperature with height. We again find transitions between energy-balanced and non-balanced situations. Interestingly, the corresponding curves do not show the strong dependence on the value of $\Lambda$ found in Fig.~\ref{figCHlambda}. Only for values of $\Lambda$ in the range from $-10h$ to $-h$ are the curves inside the limits of the domain of the plot. For the case $\Lambda=-h,$ we find that in addition to the minimum located at $x=0$ around $z=0.25h$, there are two tiny additional symmetric zones of non-equilibrium at $z=0$. These two zones are produced because conduction is negative and higher (in absolute value) than the radiative losses. For the remaining curves, the location of the height at $x=0$ where we find non-equilibrium is closely related to the condition given by Eq.~(\ref{eq:EcCHlam}).

The same analysis was performed for the AR model. The different terms in the energy equation are plotted in the three panels of Fig.~\ref{figenergytermsAR} for a temperature profile that decreases with height. The conduction term is quite low, except at the core of the AR, where it becomes negative. It is always negative when the magnetic field is purely horizontal, as Eq.~(\ref{eq:EcArlam}) indicates.  When added to the radiation term, the obtained heating distribution shows some forbidden zones represented with continuous red lines. Interestingly, the estimated heating is localised near the footpoints where the vertical component of the magnetic field is large and above the forbidden zone at the core of the AR. 

\begin{figure}[h]
\center
{\includegraphics[width=7.cm]{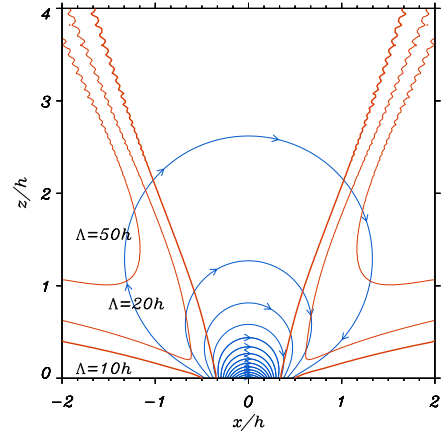}
}
\caption{\small Same as in Fig.~\ref{figCHlambda}, but for the AR model. The temperature increases with height ($\Lambda>0$).} \label{figARlambda0}
\end{figure}

\begin{figure}[h]
\center
{\includegraphics[width=7.cm]{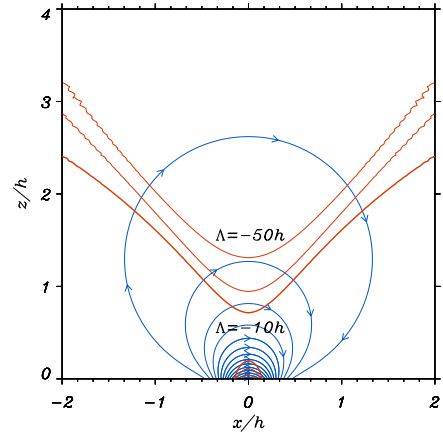}
}
\caption{\small Same as in Fig.~\ref{figARlambda0}, but now the temperature decreases with height ($\Lambda<0$). The situation shown in Fig.~\ref{figenergytermsAR} is also represented here. } \label{figARlambda1}
\end{figure}

\begin{figure}[h]
\center
{\includegraphics[width=8.cm]{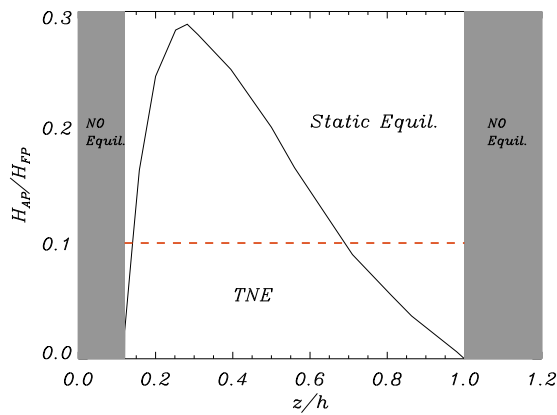}
}
\caption{\small Heating at the apex over the heating at the footpoint of a given field line as a function of height at $x=0$ (continuous line). The results shown in Fig.~\ref{figenergytermsAR} have been used to construct this plot, where the criteria of \citet{klimchukluna2019} regarding TNE have been applied ($H_{\rm AP}/H_{\rm FP}<0.1$ leads to TNE, represented with a horizontal dashed red line).} \label{figklimluna}
\end{figure}

In contrast, for a temperature that increases with height, see Fig.~\ref{figARlambda0}, the central part of the AR allows energy-balanced solutions because the conduction term is always positive at $x=0$ according to Eq.~(\ref{eq:EcArlam}). In this regard, low-lying magnetic arches are the most suitable to be in energy balance for $\Lambda>0$. However, when the temperature decreases with height, see Fig.~\ref{figARlambda1}, the curves are similar to the curve in the CH model, except that very low-lying curved fields situated at the core of the AR are now out of thermal equilibrium, as expected from the results shown in Fig.~\ref{figenergytermsAR}. These results demonstrate that small temperature gradients with height can have a significant effect on the energy equilibrium of the system, producing zones in which balance cannot be achieved. Along this line, we explored different dependences of the function $\mathcal{H}(z),$  which deviate from the simple exponential form of Eq.~(\ref{eq:h}), and we were unable to find a function that allows energy balance in the whole spatial domain. This result suggests that such a global thermal balance is quite difficult to achieve in a real magnetic structure. Non-thermal equilibrium seems to be difficult to avoid, and this might partially explain the occurrence of phenomena like coronal rain, which is regularly observed (see e.g. \citealt{antolinrouppe2012} and \citealt{antolin2020} and references therein). Nevertheless, very specific conditions are required for TNE to occur, as we describe in the following.

Finally, it is worth analysing the possible occurrence of TNE in our system using known results regarding this process. \citet{klimchukluna2019} used several approximations to derive some basic conditions that lead to the presence of TNE in coronal magnetic flux tubes. According to their results, when the ratio of the heating at the apex of a given magnetic field line over the heating at the base of the corona on the same line ($H_{\rm AP}/H_{\rm FP}$) is smaller than 0.1, the system is prone to develop cycles associated with TNE. Asymmetries can alter the conditions for TNE \citep[see also][]{pelouzeetal2021}, but they do not play any role in our perfectly symmetric configuration. We calculated this ratio from the example shown in Fig.~\ref{figenergytermsAR} using our computed heating (based on energy balance), and the results are represented in  
Fig.~\ref{figklimluna} as  a function of the height of the different field lines at $x=0$. There are two forbidden regions for heights below 0.1 and for heights above 1 (as Fig.~\ref{figenergytermsAR} already shows) in which equilibrium is not permitted. Close to these two regions, we obtain values for the heating ratios below 0.1, meaning that the system would be in TNE according to the criteria of \citet{klimchukluna2019}. Conversely, in the range of heights between $0.15\le z/h \le 0.7,$ the system would achieve static equilibrium. Although our model does not include a chromosphere, which is an elementary ingredient for TNE with periodic cycles, the calculated heating in our model can be used as a guide to infer the possible dynamical evolution of the system.

\section{Conclusions and discussion}

This study represents an exploratory first attempt at understanding the
physics of CHs and ARs from a global point of view, instead of focusing on individual magnetic field lines.  We have developed a rather
flexible, robust method for generating 2D MHS
equilibria in Cartesian geometry in the presence of constant gravity. We artificially built a magnetic field distribution at the base of the corona based on the superposition of parabolic magnetic field profiles, which were translated in terms of the flux function, $A$.  This function satisfies a Grad-Shafranov type of partial differential equation in 2D that contains a non-linear term due to the coupling with gas pressure and temperature. The magnetic field arrangement, chosen to represent open and closed magnetic field lines, was incorporated through the boundary conditions needed to solve the partial differential equation. We were able to find manageable analytical expressions for the magnetic distribution in the potential case. 

Based on physical grounds and the information provided by  observations, we proposed a relatively simple functional form for plasma pressure and temperature in terms of the flux function. We selected a depression in pressure and temperature inside the CH to have a realistic model. Under such conditions density is found to be lower inside the CH with respect to the coronal environment if the elementary constraint $p_{\rm CH}/p_{\rm C}< T_{\rm CH}/T_{\rm C}$ is satisfied. A decrease in gas pressure produces an increase in magnetic pressure  in order to keep the total pressure constant across the field lines. Along the magnetic field lines, there is, by construction in the model, balance between the gas pressure gradient and the projected gravitational term. Conversely, in our ARs models, the core is over-dense when $p_{\rm AR}/p_{\rm C}> T_{\rm AR}/T_{\rm C}$. Interestingly, the same functional form of gas pressure and temperature with the flux function correctly describes the general features of  both CHs (low pressure, temperature, and density) and ARs (high pressure, temperature, and density), which are in principle two considerably different coronal structures.  A compelling contribution of our model is that we  found solutions that can concurrently represent two different magnetic structures with properties similar to those reported by the observations. In this regard, it is imperative to mention that the exact dependence  of pressure and temperature on the flux function $A$ we considered is not a crucial step of our analysis. Other choices can also lead to physically acceptable solutions.

The highly non-linear term that couples the gas pressure effect to the magnetic field substantially complicates the derivation of analytical solutions. Nevertheless, the linearisation of the Grad-Shafranov equation leads to a Poisson equation with a source term that depends on the potential solution that is obtained from a Laplace equation. The formal solutions to these two equations were written using Green's functions and applying the method of images. The outcome of our study is a semi-analytical method, valid in the low- $\beta$ regime, that includes the effect of gas pressure and gravity. Other alternatives can be explored in the future to solve the characteristic non-linear equation without assuming that the plasma-$\beta$ is a small parameter, for example \citet{liao2003} for a novel method based on homotopy analysis. An interesting result obtained by applying our semi-analytic method is that in CHs, the density distribution is not necessarily aligned with the magnetic field. Thus, the usual assumption that density structures delineate the magnetic field in the corona is not fully justified in open structures such as CHs.

The focus of this work was finding MHS solutions under force balance, and the possible energy balance was calculated when the solution is known. We showed by calculating the conduction and radiation terms that the corresponding heating necessary to maintain energy equilibrium is constrained in the spatial domain. We also refer to \citet{petrieetal2003}, where a similar approach was used, but based on a fully analytic model. These authors where able to calculate models whose loop length, shape, plasma density, temperature, and velocity profiles were fitted to loops observed with different satellites. In our model, depending on the parameters, especially the temperature dependence on height, thermal balance may not be achieved at certain locations. Under these circumstances, the system will most likely evolve either towards a different equilibrium configuration or develop cycles around a non-existing equilibrium, such as happens in TNE processes.  In the first case, the excess of heat can be used to increase the temperature and gas pressure of the plasma along the field line. The evolution of the system strongly depends on the boundary conditions applied at the footpoints, and especially on the condition on the heat energy flux. The presence of a chromosphere at the footpoints also affects the evolution of the system, and it is especially relevant regarding TNE processes. The temporal evolution is a topic of its own and is beyond the scope of the present work, but the results presented here provide indications about the possible progression of the system with time. Because TNE is a non-linear mechanism, the absence of equilibrium in a stationary configuration does not directly imply a TNE state. Recently, several authors have carried out a parameter space investigation (see \citealt{fromentetal2018} and \citealt{pelouzeetal2021}) for TNE onset in a 1D geometry considering the possibility of asymmetric heating (along the loop) and asymmetric geometry. In addition to TNE, they also showed the conditions in which steady or static state is achieved. Importantly, they found that when the heating is small (but typical of AR), TNE occurs only when geometries and heating profiles match specific criteria. That is, at low energies, the TNE parameter space seems to be strongly constrained \citep[see also][]{klimchukluna2019}. As an illustrative example, we calculated the criteria for one of our models.

The main critical point of the approach used in this work regarding the energy balance is that the energy equation is not treated self-consistently.  Nevertheless, the exact dependence of the heating function is required to solve the problem in a self-consistent manner. The heating function has several parametrisations depending on the heating mechanism invoked (e.g. \citealt{mandrinietal2000} and \citealt{demoulinetal2003}), but the exact form of the heating function in the corona is still debated. For this reason, determining the heating function based on the constrains of force balance and thermal balance is an alternative path that should not necessarily be discarded. Although the condition of thermal or energy balance in the solar corona is not fully justified because some observations suggest the presence of low-frequency nanoflare events in AR, leading to non-steady heating, investigating the steady situation still provides  useful information. In particular, we showed that if temperature has a monotonic dependence on height (either increasing or decreasing), then we obtain regions in the magnetic configuration in which an appropriate amount of heating leads to a perfect energy balance. However, we also find regions where energy balance is not possible, and this depends on the scale of the temperature variation with height that eventually modifies the conduction and the radiative losses in the system. It might be worthwhile in the near future to compare the heating function we obtain from our models with those commonly used in the literature in order to assess the possible similarities or discrepancies among them.

%In fact, it may be possible in this way to discover systematic dependencies %among the scaling parameters.

We have used a magnetic unipolar configuration for the CH. Although there are indications in the observations that the field is unipolar based on the estimates  of  the unsigned flux, the magnetograms show the typical salt-and-pepper distribution inside CHs.  Our models only explain the upper part of the magnetic field in the solar corona, where the magnetic field is essentially open \citep[e.g.][]{wiegelmannsolanki2004}. The models we developed are fairly basic and very idealised, but contain the primitive ingredients for providing a common physical background that describes the elementary features of CHs and ARs.  Nevertheless, previous assumptions are mandatory to make analytical progress and to understand the basic physical concepts that underlie this problem. The models developed here can be extended further. In particular, a stimulating extension of our work is the analysis of MHS equilibria that contain both a CH and an AR separated by a certain distance. By analysing how the main parameters  of one structure depend on the parameters of the other, information about the coupling between the two magnetic arrangements may be gained. This may pave the path to viewing CHs and ARs as connected structures.

We have assumed that the temperature profile only depends on the magnetic field line (on the flux function $A$) and explicitly depends on height  in the coronal part of the atmosphere. Nevertheless, the model can be extended to include the lower parts of the solar atmosphere by choosing a temperature and pressure variation with height that is representative of the chromosphere, the transition region, and finally the corona. In this future model, the coupling of the corona with the chromosphere will be included and can have relevant effects on the coronal part of the solution. For this reason, it deserves to be investigated further. The thermal balance under these conditions is also of interest, but is much more complicated because of the intricate physics of the solar chromosphere. For example, the radiative losses are no longer optically thin, and therefore the loss functions used for the coronal part are not applicable. A chromospheric layer is known to act as a reservoir of energy and mass that can lead to the appearance of condensations in the corona and to cycles of thermal non-equilibrium. 

Our results should be also expanded to include flows. This is a more realistic situation than the static case because there is clear evidence in the observations of outgoing flows in CHs that are inevitably linked to the solar wind. For simplicity, we should first assume that the flow is field aligned. This problem has been investigated in the past by several authors in other contexts, see for example \citet{tsinganos1981,tsinganos1982}, \citet{lowtsinganos1986}, and \citet{webbetal1994,webbetal2001}, and in coronal loops by \citet{petrieetal2002,petrieetal2003}. When flows are present, the Grad-Shafranov equation and the Bernoulli equation are coupled. This problem is left for future studies and is also closely related to the appearance of TNE cycles. TNE is due to the necessity of an enthalpy flux to balance energy loss in the corona. In the long run, this is an unstable configuration because mass and runaway radiative losses build up.

Finally, the solutions we obtained should be extended to the 3D. This case is more
representative of a real situation. The CH model can be translated into
cylindrical geometry, while a different approach is required to represent bipolar
ARs in 3D. The logical next step is to use Euler potentials, although they have
some limitations that need to be assessed depending on the specific configuration
\citep[see details in][]{neukirch2015}. Other ad hoc approaches to building
analytic solutions in 3D that can be useful for our purposes have been
investigated by \citet{low1985,low1991}, \citet{neukirch1999}, and
\citet{neukirch2019}, although a purely numerical treatment is most likely
required in 3D when the thermal structure is included in the problem, as in our
study here.

\begin{acknowledgements}
This publication is part of the R+D+i project PID2020-112791GB-I00, financed by MCIN/AEI/10.13039/501100011033. M. L. acknowledges support through the Ram\'on y Cajal
fellowship RYC2018-026129-I from the Spanish Ministry of
Science and Innovation, the Spanish National Research Agency
(Agencia Estatal de Investigaci\'on), the European Social Fund
through Operational Program FSE 2014 of Employment,
Education and Training and the Universitat de les Illes Balears.
I. A. was supported by project PGC2018-102108-B-I00 from Ministerio de Ciencia, Innovaci\'on y Universidades and FEDER funds. P. A. acknowledges funding from his STFC Ernest Rutherford Fellowship (No. ST/R004285/2). This research was funded in part by the Austrian Science Fund (FWF): J4624-N. The authors thank the referee, Prof. Tsinganos, for useful comments that helped to improve the paper.
\end{acknowledgements}

%\bibliographystyle{apj}      % basic style, author-year citations
%\bibliography{jaume}   % name your BibTeX data base

\bibliography{main.bbl}

\end{document}